\documentclass[preprint,3p,times,onecolumn]{elsarticle}

\usepackage{graphicx,amsmath,times,hhline}
\usepackage{amssymb,graphicx}
\usepackage{amsmath,bm}
\usepackage{fancyhdr}

\usepackage{hyperref}
\biboptions{numbers,sort&compress}

\newcommand{\bea}{\begin{eqnarray}}
\newcommand{\eea}{\end{eqnarray}}
\newcommand{\bes}{\begin{subequations}}
\newcommand{\ees}{\end{subequations}}
\newcommand{\ds}{\displaystyle}

\def\R{\hbox{{\rm I}\kern-0.2em{\rm R}\kern0.2em}}%mathematical R for reals
\def\a{\alpha}  
\def\b{\beta}          \def\g{\gamma}

\def\a{\alpha}
\def\b{\beta}
\def\g{\gamma}

\def\bn{\begin{equation}}\def\en{\end{equation}}
\def\bny{\begin{eqnarray}}\def\eny{\end{eqnarray}}
\def\be{\begin{eqnarray*}}\def\ee{\end{eqnarray*}}

\def\p{\partial}
\def\({\left(}
\def\){\right)}
\def\[{\left[}
\def\]{\right]}
\def\bc{\begin{center}}
\def\ec{\end{center}}

\journal{Applied Mathematics and Computation}

\begin{document}

\begin{frontmatter}
\title{Lie symmetry analysis and group invariant solutions of the\\ nonlinear Helmholtz equation}

\author[ks]{K. Sakkaravarthi}\ead{ksakkaravarthi@gmail.com}
\author[agp]{A.G. Johnpillai}\ead{andrewgratienj@yahoo.com}
\author[ad]{A. Durga Devi}
\author[tk]{T. Kanna}\ead{kanna{\_}phy@bhc.edu.in}
\author[ks]{M. Lakshmanan\corref{ml}}\ead{lakshman.cnld@gmail.com; lakshman@cnld.bdu.ac.in}

\address[ks]{Centre for Nonlinear Dynamics, School of Physics, Bharathidasan University, Tiruchirappalli -- 620 024, India}
\address[agp]{Department of Mathematics, Eastern University, Chenkalady -- 30350, Sri Lanka}
\address[ad]{Department of Physics, Srinivasa Ramanujan Center, SASTRA University, Kumbakonam - 612 001, India}
\address[tk]{Post-Graduate and Research Department of Physics, Bishop Heber College, Tiruchirappalli -- 620 017, India}

\cortext[ml]{Corresponding author. Tel.: +91 431 2407093; Fax: +91 431 2407093}

\begin{abstract}
We consider the nonlinear Helmholtz (NLH) equation describing the beam propagation in a planar waveguide with Kerr-like nonlinearity under non-paraxial approximation. By applying the Lie symmetry analysis, we determine the Lie point symmetries and the corresponding symmetry reductions in the form of ordinary differential equations (ODEs) with the help of the optimal systems of one-dimensional subalgebras. Our investigation reveals an important fact that in spite of the original NLH equation being non-integrable, its symmetry reductions are of Painlev{\'e} integrable. We study the resulting sets of nonlinear ODEs analytically either by constructing the integrals of motion using the modified Prelle-Singer method or by obtaining explicit travelling wave-like solutions including solitary and symbiotic solitary wave solutions. Also, we carry out a detailed numerical analysis of the reduced equations and obtain multi-peak nonlinear wave trains. As a special case of the NLH equation, we also make a comparison between the symmetries of the present NLH system and that of the standard nonlinear Schr{\"o}dinger equation for which symmetries are long available in the literature. 
\end{abstract}

\end{frontmatter}

%\maketitle

\section{Introduction} %\setcounter{equation}{0}
Investigation of physical systems described by nonlinear evolution equations (NLEEs) and exploring their underlying dynamics remain the central focus of research for the past few decades. Finding exact solutions of these nonlinear equations, which can be either ordinary differential equations (ODEs) or partial differential equations (PDEs), is one of the most important tasks and their further investigation plays a crucial role in the study of nonlinear physical phenomena.
Lie symmetry analysis \cite{BK,Ib,Ol} has a time honoured history and has been proved to be a powerful tool for studying nonlinear problems arising in many scientific fields including mathematics, physics, and biology.
A Lie (point) symmetry of a given system of NLEE(s) is nothing but an infinitesimal transformation of all of its (their) independent and dependent variables, which leaves the corresponding nonlinear equations invariant under
that transformation. From the Lie point symmetries, with the aid of the invariance conditions, a given system of equations is reduced to a set of ODEs/PDEs with lesser number of independent coordinates.
The reduced equation becomes an (or a set of) ODE(s) if the independent variables involved in the original NLEE are two, while it becomes a (or a set of) PDE(s) when the NLEE contains more than two independent variables. Thus a similarity reduction of a differential equation is closely connected with the invariance of the respective equation. For a detailed background on the theory and application of the Lie symmetry methods one can refer to Refs. \cite{BK,Ib,Ol,Ov,PK-ML,PK-ML2,Ol2,winter,nls-res}.

This work deals with the Lie symmetry analysis and the construction of group invariant solutions of a physically important evolution equation, namely the following nonlinear Helmholtz (NLH) equation
\bea
iq_t+k~q_{tt}+q_{xx} +\gamma~|q|^2q=0, \label{0}
\eea
where $q$ is a complex valued dependent variable representing the envelope field, the independent variables $t$ and $x$ denote the longitudinal and transverse coordinates, respectively. Here $k$ and $\gamma$ are arbitrary real constants and represent the coefficients of non-paraxial term and the cubic nonlinearity, respectively.
Equation (\ref{0}) is receiving much attention in different areas of physics and mathematics. 
In the context of nonlinear optics, Eq. (\ref{0}) describes the ultra-broad beam propagation in a Kerr-like nonlinear medium under the non-paraxial approximation \cite{crosi}. Due to this reason, Eq. (\ref{0}) can also be termed as non-paraxial nonlinear Schr\"odinger equation \cite{McD}. Physically, Eq. (\ref{0}) can be viewed as an optical system governing the propagation of beam in a Kerr type nonlinear medium experiencing diffraction in both the  transverse and longitudinal directions. The NLH equation (\ref{0}) seems to be non-integrable as it does not posseses a Lax pair. It also does not admit sufficient number of conserved quantities. So far to the best of our knowledge, there is no claim for the integrability of the NLH equation (\ref{0}) but it admits interesting special solutions. Depending upon the sign of the nonlinearity coefficient $\g$, Eq. (\ref{0}) reduces to a focusing-type NLH equation (for $\g>0$) or a defocusing-type NLH equation (for $\g<0$), for which respectively bright and dark solitary-wave solutions have been reported  \cite{McD2,McD3}. 
In recent years the multicomponent NLH system has also received considerable interest. Special solitary waves and elliptic waves for a multicomponent NLH system exhibiting distinct dynamical behaviours have been reported in \cite{TK-nlh}.

On the other hand, under the paraxial approximation the Maxwell's equations result in the following standard nonlinear Schr\"odinger (NLS) equation \cite{nls-ref}:
\bea
iq_t+q_{xx} +\gamma~|q|^2q=0. \label{nls}
\eea
Mathematically, NLS equation (\ref{nls}) can be obtained from the NLH equation (\ref{0}) as a special case, for $k=0$. Physically, it can be derived to describe a beam propagating in a Kerr type nonlinear medium, where the beam
radius is sufficiently large compared with the beam wavelength which supports only the paraxial approximation \cite{nls-ref}. Here the second order derivative of the field variable with respect to propagation direction
($q_{tt}$) vanishes due to the exclusion of non-paraxial effect ($k=0$). The complete integrability of NLS equation (\ref{nls}) has been estabilished via the inverse scattering transformation method, the Painlev\'e analysis, admittance of infinite number of conserved quantities, etc. \cite{nls-int}.
Also, the NLS equation admits multi-soliton solutions (bright solitons for $\g>0$ and dark solitons for $\g<0$) in addition to several other exact analytical solutions and display elastic interaction of solitons \cite{nls-int}.% \cite{mkj}.

The objectives of the present work are three-fold. First and the important objective is to obtain the symmetry reductions of the NLH system (\ref{0}) using the infinitesimal similarity transformations. These similarity transformations are constructed by utilizing the Lie point symmetry generators admitted by the complex PDE (\ref{0}). In addition to the construction of symmetry generators, we obtain a set of associated nonlinear ODEs as the integrable reductions of the NLH system (\ref{0}). Second, we investigate the integrable nature of the reduced nonlinear ODEs by applying the Painlev\'e singularity structure analysis. Then, we derive the first integrals/constants of motion of these nonlinear ODEs by adopting the modified Prelle-Singer method and study the reduced ODEs by means of direct numerical analysis. The third objective is to revisit the symmetry analysis of the NLS equation (\ref{nls}) \cite{winter,nls-res} and to compare the results with that of the present NLH equation (\ref{0}).

The outline of the paper is as follows. In Section 2, we briefly discuss the general algorithm for the Lie point symmetry analysis. We present the symmetry reductions and group-invariant solutions of the NLH equation (\ref{0}) in Sec. 3. Section 4 deals the symmetry reductions of the NLS equation (\ref{nls}) and their relevance to the NLH system. Finally, in Section 5, we summarize the main results of the present work.

\section{General algorithm for Lie symmetries and invariants}
In this section, we briefly explain the important steps involved in the determination of Lie symmetries for a given system of (complex or real) nonlinear PDEs \cite{BK,Ib,Ol,Ov,PK-ML,PK-ML2,Ol2,winter,nls-res}. Firstly, the given equation has to be rewritten as a function of all dependent and independent variables by using a similarity transformation with an infinitesimal parameter. For example, the nonlinear equation (having $N$ dependent functions $u_j$ and two independent functions $t$ and $x$) of the form
\bea
{F}_j(u_1,u_2,u_3,...,u_N,x,t)=0,\qquad j=1,2,3,...,N, \label{pde}
\eea
can be transformed to a set of PDEs using the following infinitesimal transformations:
\bes\bny
&&{u}_j \rightarrow \hat{u}_j = u_j+\epsilon~ \eta^j(x,t,u_1,u_2,u_3,...,u_N)+O(\epsilon^2), \qquad j=1,2,3,\cdots ,N,\\
&&~~{t} \rightarrow ~\hat{t}=t+\epsilon~ \tau(x,t,u_1,u_2,u_3,...,u_N)+O(\epsilon^2),\\
&&~~{x} \rightarrow \hat{x}= x+\epsilon~ \xi(x,t,u_1,u_2,u_3,...,u_N)+O(\epsilon^2),
\label{trans}\eny\ees
where $\epsilon$ is a small expansion parameter. According to the Lie symmetry algorithm, a vector field \bn
X=\tau(t,x,u_1,u_2,...,u_N)\p_t+\xi(t,x,u_1,u_2,...,u_N)\p_x+\ds\sum_{j=1}^N \eta^j(t,x,u_1,u_2,...,u_N)\p_{u_j}, \label{2}\en
is a generator of point symmetry of the equation
(\ref{pde}) if \bny
X^{[2]} ~{F}_j(u_1,u_2,u_3,...,u_N,x,t)\big|_{F_j=0}&=&0.\label{3}\eny
Here the operator $X^{[2]}$ is the second prolongation of the
operator $X$ which is determined by the order of the given differential
equation and is defined as
\bny X^{[2]}=X+\ds\sum_{j=1}^N\zeta_t^j \p_{u_{j,t}}+\ds\sum_{j=1}^N\zeta_x^j\p_{u_{j,x}}+\ds\sum_{j=1}^N\zeta_{xx}^j\p_{u_{j,xx}}, \eny
where the coefficients $\zeta^j_t$, $\zeta^j_x$, and $\zeta^j_{xx} $ correspond to the  prolongation formulae, $\zeta_t^j=D_t(\eta^j)-u_{j,t}D_t(\tau)-u_{j,x}D_t(\xi)$, $\zeta_x^j=D_x(\eta^j)-u_{j,t}D_x(\tau)-u_{j,x}D_x(\xi)$, $\zeta_{xx}^j=D_x(\zeta_x^j)-u_{j,xt}D_x(\tau)-u_{j,xx}D_x(\xi)$.
%\bes\bny \zeta_t^j&=&D_t(\eta^j)-u_{j,t}D_t(\tau)-u_{j,x}D_t(\xi),\\
%\zeta_x^j&=&D_x(\eta^j)-u_{j,t}D_x(\tau)-u_{j,x}D_x(\xi),\\
%\zeta_{xx}^j&=&D_x(\zeta_x^j)-u_{j,xt}D_x(\tau)-u_{j,xx}D_x(\xi). \eny\ees 
Here the index `$j$' ($j=1,2,3,...,N$) denotes the dependent variable and the subscripts represent the independent variables $t$ and $x$, while $D_t$ and $D_x$ are the standard total derivative operators.
%defined by \bn D_t=\p_t +\ds\sum_{j=1}^N u_{j,t}\p_{u_j}, \qquad D_x=\p_x+\ds\sum_{j=1}^N u_{j,x}\p_{u_j}.\en

One can obtain the explicit form of the infinitesimal
coefficients $\tau, \xi$ and $\eta^j$ by solving the determining
equations (\ref{3}) with required number of arbitrary constants
through tedious but straightforward calculations.
From these infinitesimal coefficients and Eq. (\ref{2}), the Lie point
symmetry generators (vector fields) admitted by the system
of PDEs (\ref{pde}) corresponding to each arbitrary constants can be furnished in a direct way.
The symmetry reductions of the given equations can be identified by solving the following characteristic equation
\bea \frac{dt}{\tau}=\frac{dx}{\xi}=\frac{du_1}{\eta^1}=\frac{du_2}{\eta^2}=\frac{du_3}{\eta^3}=...=\frac{du_N}{\eta^N}. \eea
Here one of the important points to note is that we wish to minimize the search for invariant solutions by finding the nonequivalent branches of solutions resulting for different combinations (linear superposition) of symmetry generators.
For this purpose, an optimal system of one-dimensional subalgebras is constructed with the help of adjoint representations of the generators and this gives exact invariants as well as the associated symmetry reductions.
Specifically, the vector fields ($X_j$) are written as a superposition, with equal number of arbitrary coefficients ($c_j$), in the form $X=c_1 X_1+c_2 X_2+c_3 X_3+\cdots +c_n X_n \equiv \sum_{j=1}^n c_j X_j$, 
%\bny X=c_1 X_1+c_2 X_2+c_3 X_3+\cdots +c_n X_n \equiv \ds\sum_{j=1}^n c_j X_j, \label{sup} \eny
where $j=1,2,3,\cdots,n$, and $n$ is the number of available vector fields. Here one has to reduce this equation to the simplest forms for different choices of the coefficients $c_j$ by using the adjoint characteristics of the vector fields defined as $\mbox{Ad}(\varepsilon X_i,X_j)=X_j-\varepsilon [X_i,X_j]+\frac{\varepsilon^2}{2!}[X_i,[X_i,X_j]] -\frac{\varepsilon^3}{3!}[X_i,[X_i,[X_i,X_j]]]+ \ldots,$
%\bny \mbox{Ad}(\varepsilon X_i,X_j)=X_j-\varepsilon [X_i,X_j]+\frac{\varepsilon^2}{2!}[X_i,[X_i,X_j]] -\frac{\varepsilon^3}{3!}[X_i,[X_i,[X_i,X_j]]]+ \ldots,\quad  \label{adj}\eny 
where the square bracket $[X_i,X_j]$ represents the usual commutator.
For more details regarding the one-dimensional subalgebras, see \cite{Ol,readiff} and references therein. Especially, such type of group classification for certain systems of nonlinear reaction diffusion equations have been investigated in \cite{readiff}. Thus for each optimal subset of the generators we obtain appropriate invariants which reduce the original PDEs to a set of ODEs. Then we have to analyze these reductions and solve them to find their explicit solutions which can reveal the real dynamics of the physical problem of interest. By following the above algorithm, in the following sections, we perform the Lie point symmetry analysis of the NLH equation and also compare the results with those of the NLS equation.

\section{Lie symmetry analysis of the nonlinear Helmholtz equation (\ref{0})}%\setcounter{equation}{0}
In this section, we systematically derive the Lie point symmetries of the NLH equation (\ref{0}). For this purpose, we denote $q(x, t)=u(x, t)+i v(x, t)$, and decompose equation (\ref{0}) into real and imaginary parts to obtain the following system of partial differential equations (PDEs):
\bes\bny u_t+kv_{tt}+v_{xx}+\gamma\,v(u^2+v^2)=0\equiv F_1, \\
v_t-ku_{tt}-u_{xx}-\gamma\,u(u^2+v^2)=0\equiv F_2.\eny  \label{nlh1}\ees
From the previous section, we write the symmetry transformations as ${u} \rightarrow {u}_1 = u+\epsilon~ \eta^1(x,t,u,v)$, ${v} \rightarrow {u}_2 = v+\epsilon~ \eta^2(x,t,u,v)$, ${t} \rightarrow \hat{t} = t+\epsilon~ \tau(x,t,u,v)$, and ${x} \rightarrow \hat{x} = x+\epsilon~ \xi(x,t,u,v)$. Then the generalized vector field associated with Eq. (\ref{nlh1}) can be written as
\bn X=\tau(t,x,u,v)\p_t+\xi(t,x,u,v)\p_x+\eta^1(t,x,u,v)\p_u+\eta^2(t,x,u,v)\p_v . \label{nlh2}\en
The Lie symmetry/ invariance conditions of the PDEs (\ref{nlh1}) read
\bes\bny
X^{[2]}\left[u_t+kv_{tt}+v_{xx}+\gamma\,v(u^2+v^2)\right]\big|_{F_1=0}=0,\\
X^{[2]}\left[v_t-ku_{tt}-u_{xx}-\gamma\,u(u^2+v^2)\right]\big|_{F_2=0}=0,\eny \label{nlh3}\ees
where $X^{[2]}$ denotes the second prolongation operator and it takes the form (\ref{3}) with $N=2$.

We have obtained the coefficient functions $\tau, \xi, \eta^1$ and $\eta^2$
by solving the determining equations resulting from (\ref{nlh3}).
Particularly, we find that $\tau, \xi, \eta^1$ and $\eta^2$ are independent of the
derivatives of $u$ and $v$. So, the coefficients of like derivatives
of $u$ and $v$ in Eq. (\ref{nlh3}) are equated to yield an over
determined system of linear PDEs. Solving these equations recursively,
we have obtained the form of the infinitesimal coefficients
$\tau, \xi, \eta^1$ and $\eta^2$ of the NLH equation (\ref{nlh1}) as
\bes\bny
\tau=c_1+2kc_4 x, &\qquad &
\xi=c_2-2c_4 t , \\
\eta^1=(c_4 x-c_3)v ,&\qquad &
\eta^2=-(c_4 x-c_3)u,
\eny \label{ic1} \ees
where $c_1$, $c_2$, $c_3$ and $c_4$ are arbitrary real constants.
Thus, the explicit expression of the generalized vector field (\ref{nlh2}) can be written as
\bn X=(c_1+2kc_4 x)\p_t+(c_2-2c_4 t)\p_x+(c_4 x-c_3)v\p_u-(c_4 x-c_3)u(t,x,u,v)\p_v . \en
Also, from the explicit form of the infinitesimal coefficients (\ref{ic1}), the Lie point
symmetry generators for the four arbitrary constants admitted by the NLH system
of PDEs (\ref{nlh1}) are obtained as below:
\bes \bny
&X_1=\p_t,\qquad  X_2=\p_x,\qquad  X_3=-v\,\p_u+u\p_v, & \\
&X_4 =2kx\,\p_t-2t\,\p_x+xv\,\p_u-xu\,\p_v.& \eny \label{nlh4}\ees
Note that the first three Lie symmetry generators $X_1$, $X_2$, and $X_3$ can be associated with translation in time, translation in space, and phase transformations, respectively.

Now, we utilize the aforementioned Lie point symmetry generators (\ref{nlh4}) of the
NLH equation (\ref{0}) to obtain its symmetry reductions. Then we construct exact
group-invariant solutions by reducing Eq. (\ref{nlh1}) into a set of ODEs.
In this regard, we first construct the optimal system of
one-dimensional subalgebras from the generators (\ref{nlh4}).
Here the commutator and adjoint property corresponding to each
generator set plays a crucial role in determining the subalgebras.
We have given such commutator and adjoint properties of the
generators in Table 1 and Table 2, respectively.
\begin{center}%\footnotesize
\textbf{Table 1.} Commutator table of the vector fields of Eq. (\ref{nlh1})
\\[2ex]
\begin{tabular}{ccccc}\hline
$[X_i,X_j]$ &$X_1$ & $X_2$ & $X_3$ & $X_4$ \\ \hline
$X_1$ & $0$ & $0$& $0$ & $-2X_2$  \\
$X_2$ & $0$ & $0$& $0$ & $2k X_1-X_3$ \\
$X_3$ & $0$ & $0$& $0$ & $0$ \\
$X_4$ & $2X_2$ & $X_3-2k X_1$ & $0$ & $0$\\\hline
\end{tabular}
\end{center}

\begin{center}%\footnotesize
\textbf{Table 2.} Adjoint table of the vector fields of Eq. (\ref{nlh1})
\\[2ex]
\begin{tabular}{ccccc}\hline
{\rm Ad}$(\varepsilon X_i,X_j)$ &$X_1$ & $X_2$ & $X_3$ & $X_4$\\ \hline
$X_1$ & $X_1$ & $X_2$& $X_3$ & $X_4+2\varepsilon X_2$  \\
$X_2$ & $X_1$ & $X_2$& $X_3$ & $X_4-\varepsilon (2kX_1-X_3)$  \\
$X_3$ & $X_1$ & $X_2$& $X_3$ & $X_4$ \\
$X_4$ & $X_1-2\varepsilon X_2+\varepsilon^2 (X_3-2kX_1)$ & $X_2-\varepsilon (X_3-2kX_1)-2 \varepsilon^2 kX_2$& $X_3$ & $X_4$ \\
\hline
\end{tabular}
\end{center}

After a careful analysis, we find that the present NLH system
admits the following optimal set of one-dimensional subalgebras
as a linear superposition of symmetry generators (\ref{nlh4}):
(i) $X_1+aX_2$, (ii) $X_2+aX_3$, (iii) $X_1+aX_2+b X_3$ and (iv) $X_4$,
where $a$ and $b$ are arbitrary real constants.
By solving the resulting set of characteristic equations for the above subalgebras,
we get the invariants and group-invariant solutions for Eq. (\ref{nlh1}) as given in the Table 3.
\begin{center}%\footnotesize
\textbf{Table 3.} Subalgebra, group invariants, and group-invariant solutions of Eq. (\ref{nlh1})
\\[2ex]
\begin{tabular}{clcl}\hline
Case & Subalgebra & Invariant `$y$'~~~ &  {\rm Invariant~ solution} \\
\hline
(i) & $X_1+aX_2$ & $x-at$ & $u=A(y)$,  $v=B(y)$\\
(ii)& $X_2+aX_3$ & $t$ & $u=-A(y)\sin ax +B(y)\cos ax$ \\ & & & $v=A(y)\cos ax +B(y)\sin ax$\\
(iii)& $X_1+aX_2+b X_3$ & $x-at$ & $u=-A(y)\sin b t +B(y)\cos b t$ \\ & & & $v=A(y)\cos b t +B(y)\sin b t$\\
(iv)& $X_4$ & $kx^2+t^2$ & $u=A(y)\cos (\frac{t}{2k})+B(y)\sin (\frac{t}{2k})$ \\ & & & $v=-A(y)\sin (\frac{t}{2k})+B(y)\cos (\frac{t}{2k})$\\
\hline
\end{tabular}
\end{center}
In what follows, we analyze the invariant reductions obtained in the form of coupled ODEs and the associated integrals for the above four cases.
We obtain exact analytical solutions for certain equations in terms of travelling waves while the other equations are studied numerically to explore the nature of the solitons.
%In the following part, we analyze the corresponding invariant ODEs reductions and the integrals of motions. For certain equations, we present their straightforward analytical solutions. Also, we investigate these equations numerically and bring out the dynamics of periodic wave solutions.

\subsection*{{\bf Case (i)}: $X_1+aX_2$ (Travelling wave solutions)}
In this case, substitution of the group-invariant solutions (case (i) in Table 3) into the PDEs (\ref{nlh1}) results in the following system of nonlinear second-order ODEs:
\bes\bny (1+ka^2)A^{''}+aB'+\g A(A^2+B^2)=0,  \\
(1+ka^2)B^{''}-aA'+\g B(A^2+B^2)=0. \eny \label{5}\ees
Here and in the following `prime' appearing in $A$ and $B$ represents differentiation with respect to the invariant variable $y$.
In order to know about the integrable nature of the reduced ODEs, we have performed the Painlev\'e singularity structure analysis \cite{pain} and find that the above coupled second-order ODE is Painlev\'e integrable for arbitrary $a$, $k$ and $\g$ parameters. Details of the Painlev\'e
analysis to Eq. (\ref{5}) are given in \ref{sec-pain}.

Next we investigate Eq. (\ref{5}) by using the modified Prelle-Singer method \cite{PS,Duarte,ePS} (a brief algorithm for this method is given in \ref{app-ePS}), an effective method to obtain the first integrals. The explicit forms of the first integrals of Eq. (\ref{5}) obtained by this method are given below:
\bes\bny	&&I_1=2({A'^2+B'^2})+\frac{\gamma(A^2+B^2)^2}{1+k a^2},\\
&&I_2=(A'B-AB')+\frac{a(A^2+B^2)}{2(1+k a^2)},	\eny \label{5int} \ees
where $I_1$ and $I_2$ are the two first integrals of Eq. (\ref{5}). As the equation (\ref{5}) is a system of two coupled second order autonomous ODEs, the existence of two functionally independent `time-independent' integrals itself ensures the complete integrability of the system. One can also proceed to find two additional time dependent integrals to deduce the explicit solutions as well. The above integrals given by Eq. (\ref{5int}) can also be viewed as conserved quantities associated with the symmetry generators $X_1+a X_2$.
\begin{figure}[h]
\centering \includegraphics[width=0.32\linewidth]{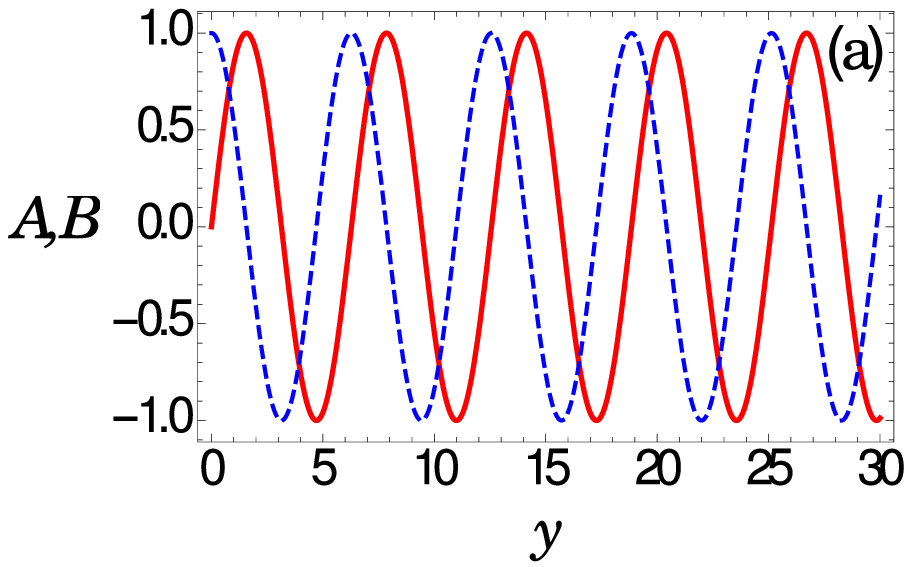} \includegraphics[width=0.32\linewidth]{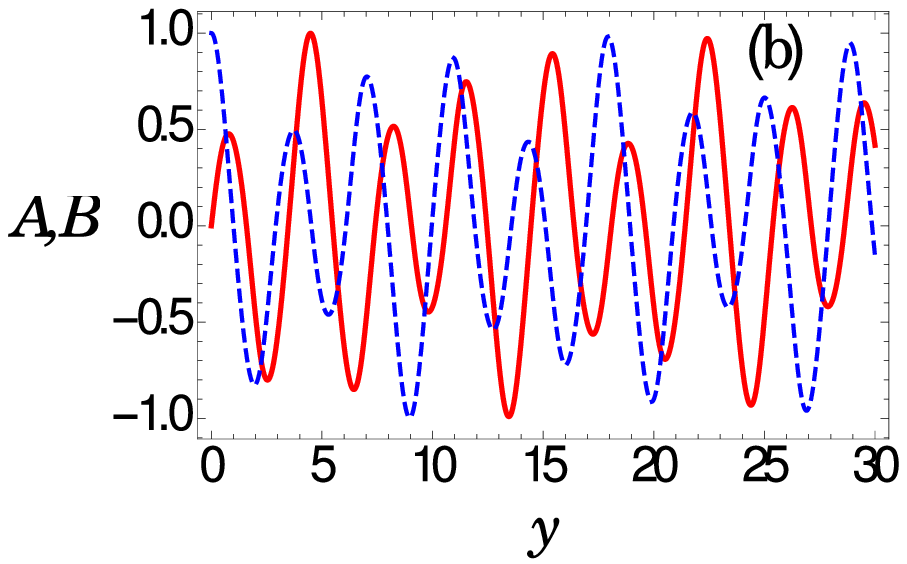} \\
\centering \includegraphics[width=0.32\linewidth]{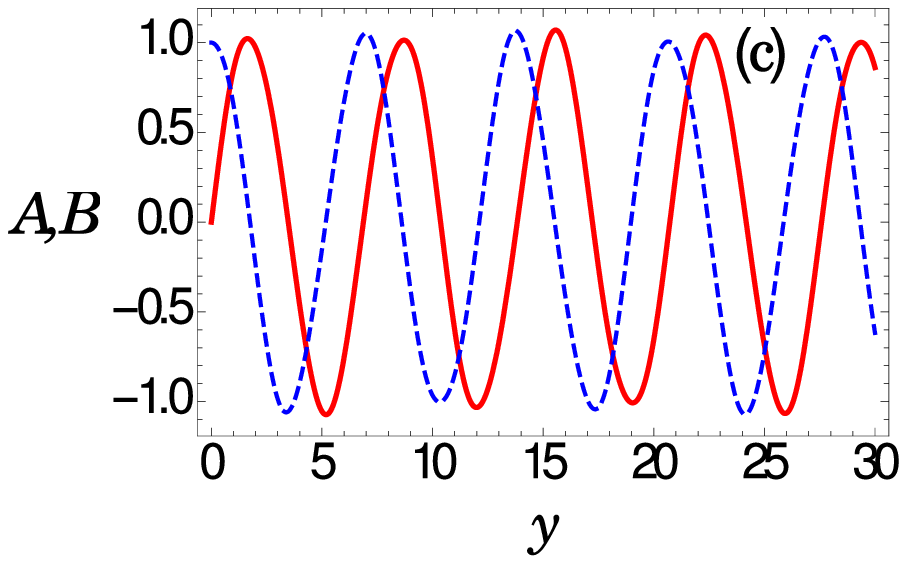} \includegraphics[width=0.32\linewidth]{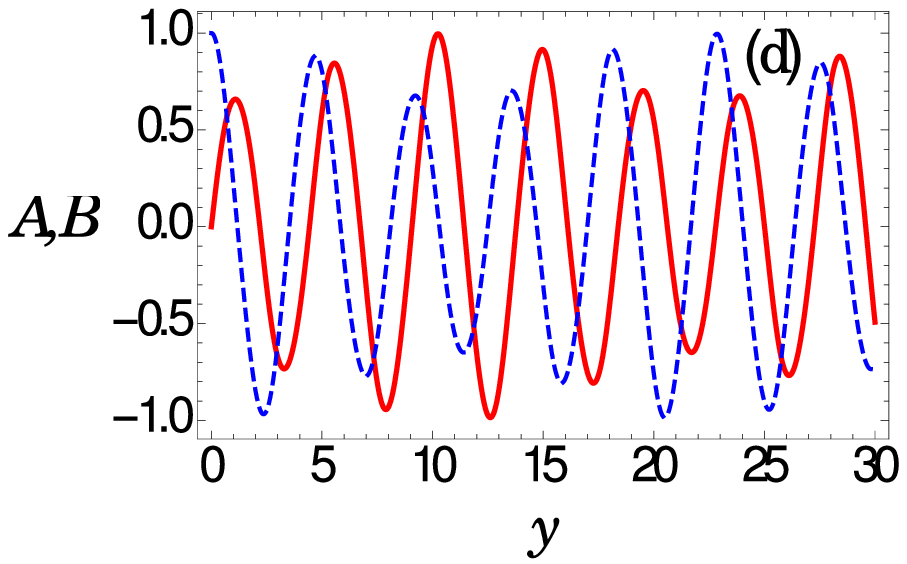} \\ \includegraphics[width=0.32\linewidth]{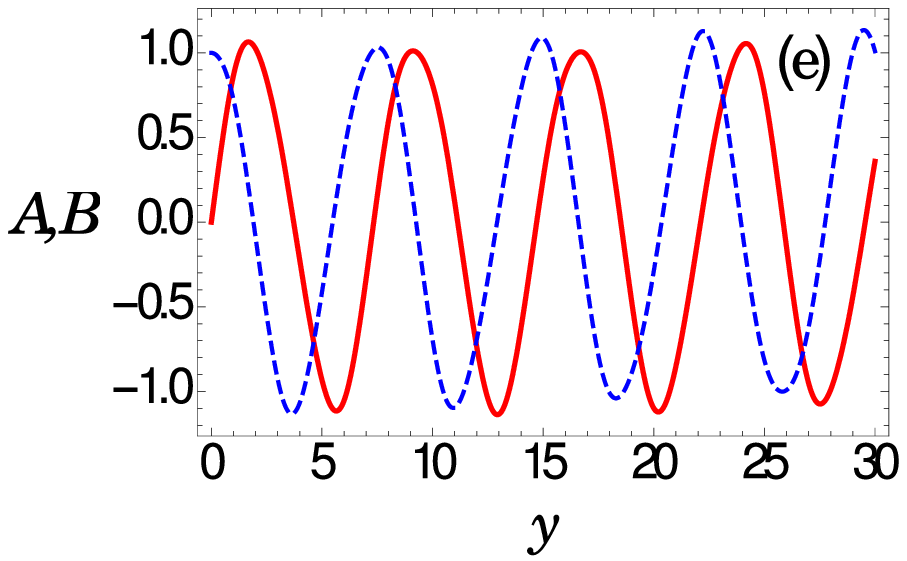}   \includegraphics[width=0.32\linewidth]{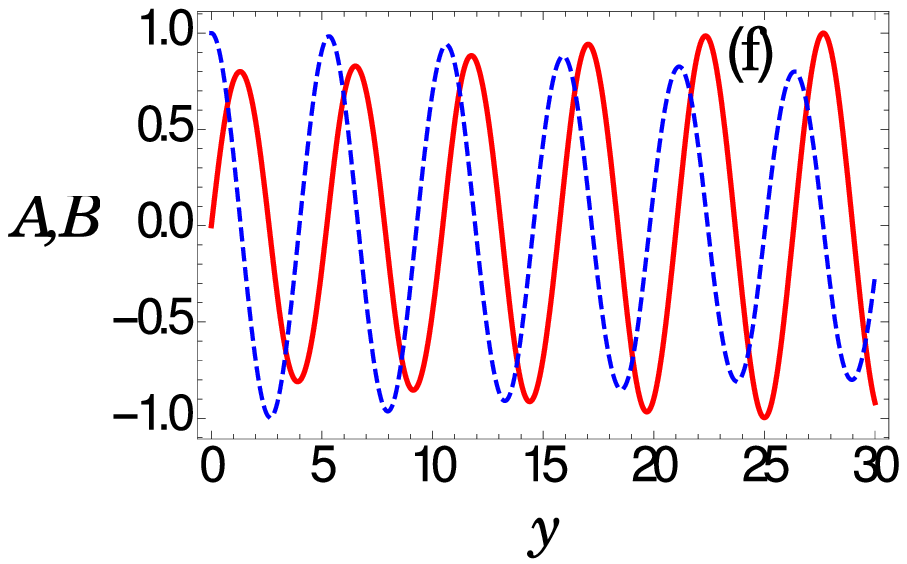}
\caption{Nonlinear periodic wave trains of Eq. (\ref{5}) for different non-paraxial coefficient $k=0,~0.5,~1$ (from top to bottom) and $\gamma=2$. The left and right panels correspond to $a=1$ and $a=-1$, respectively. Here and in the figures follow, the solid-red line represents the solution of `$A$' and the dashed-blue line denotes the solution of `$B$'.}
\label{fg1}
\end{figure}

Our numerical analysis of Eq. (\ref{5}) shows that it admits nonlinear periodic wave structures for $A(y)$ and $B(y)$. By changing the parameters $k$ and $a$ the period of oscillations and amplitudes can be altered as shown in Fig. \ref{fg1}. The top panels show the nonlinear periodic waves in the absence of non-paraxial effect ($k=0$), which corresponds to the NLS case, while the middle ($k=0.5$) and bottom ($k=1$) panels display the nonlinear periodic waves in its presence. A noticeable observation is that due to the non-paraxial effect the width of the pulse increases. Also, the nonlinear wave structures approach a stable profile with less fluctuations in their amplitude, especially when $a=-1$. One can evidence the applications of such nonlinear periodic waves in fiber Bragg gratings.

\begin{figure}[h]
\centering \includegraphics[width=0.32\linewidth]{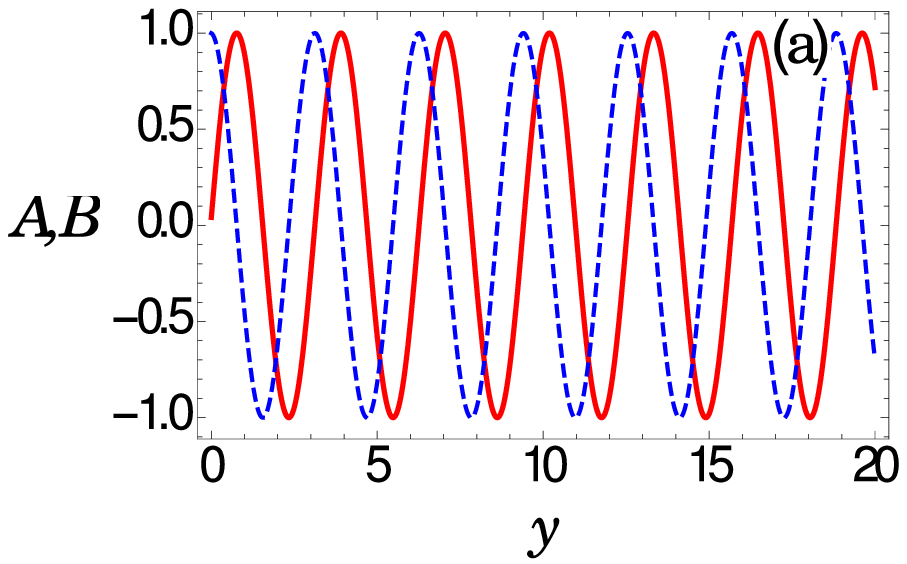} \includegraphics[width=0.32\linewidth]{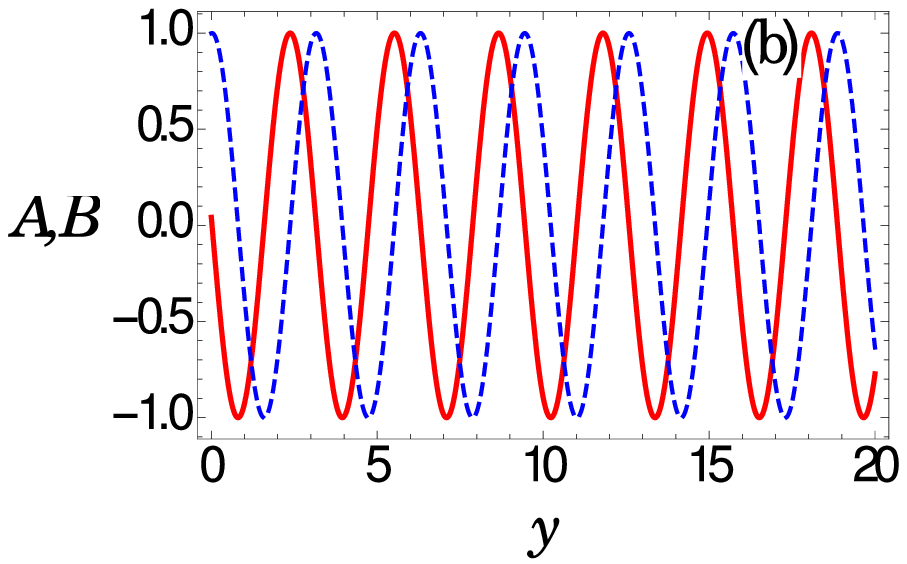}\\
\centering \includegraphics[width=0.32\linewidth]{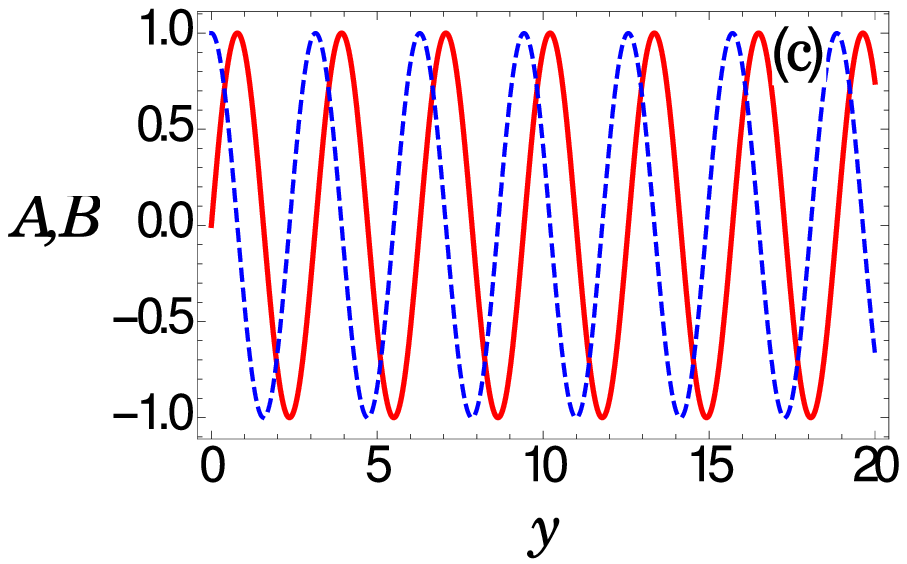} \includegraphics[width=0.32\linewidth]{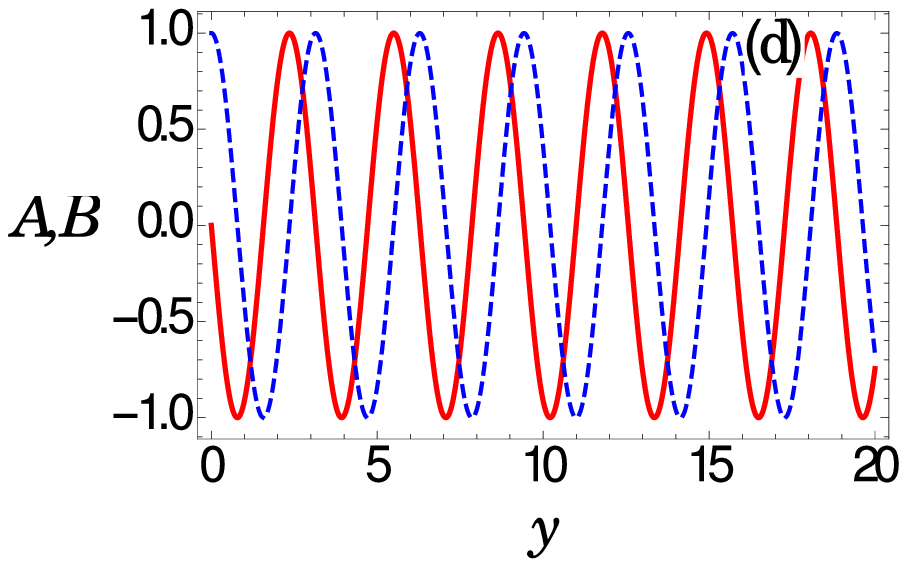}
\caption{Periodic sine and cosine wave solutions of Eq. (\ref{5a}) for $a= 1$ (a,c) and $a= -1$ (b,d) with $\gamma=2$. Here the top panel shows the analytical solution for $b_1=0.5$ and $b_2=3$, while the bottom panel represents its direct numerical solution. $A$: solid-red line and $B$: dashed-blue line.}
\label{fg2}
\end{figure}

\begin{figure}[h]
\centering \includegraphics[width=0.3\linewidth]{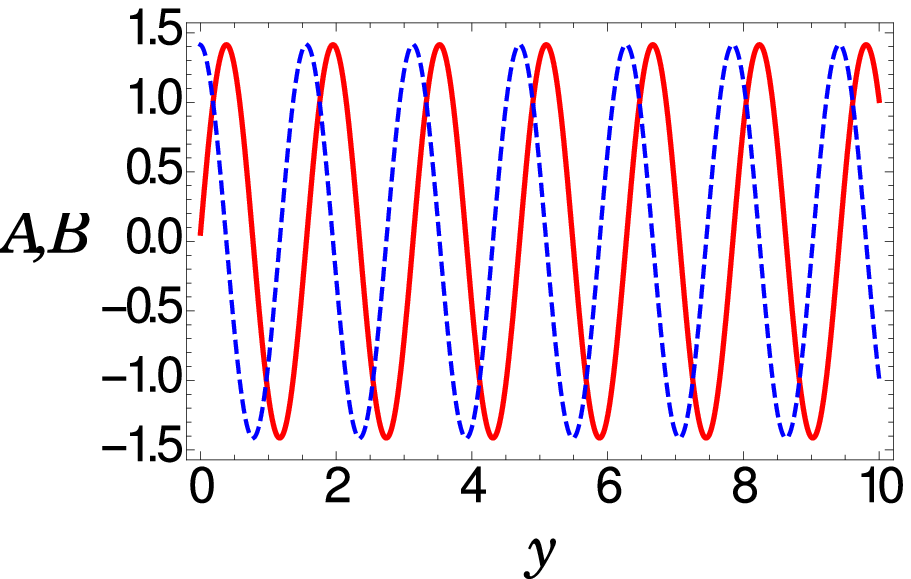}~~ \includegraphics[width=0.3\linewidth]{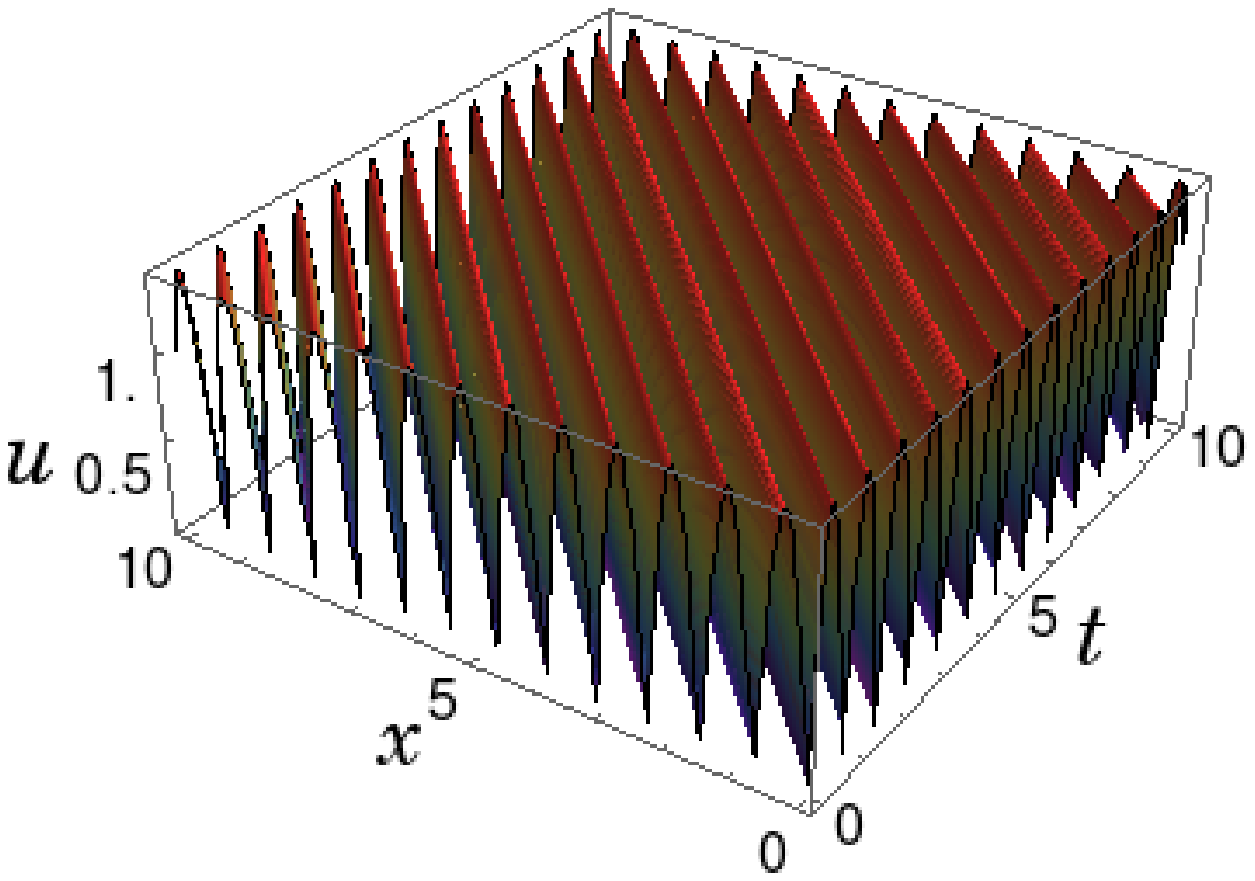} \includegraphics[width=0.3\linewidth]{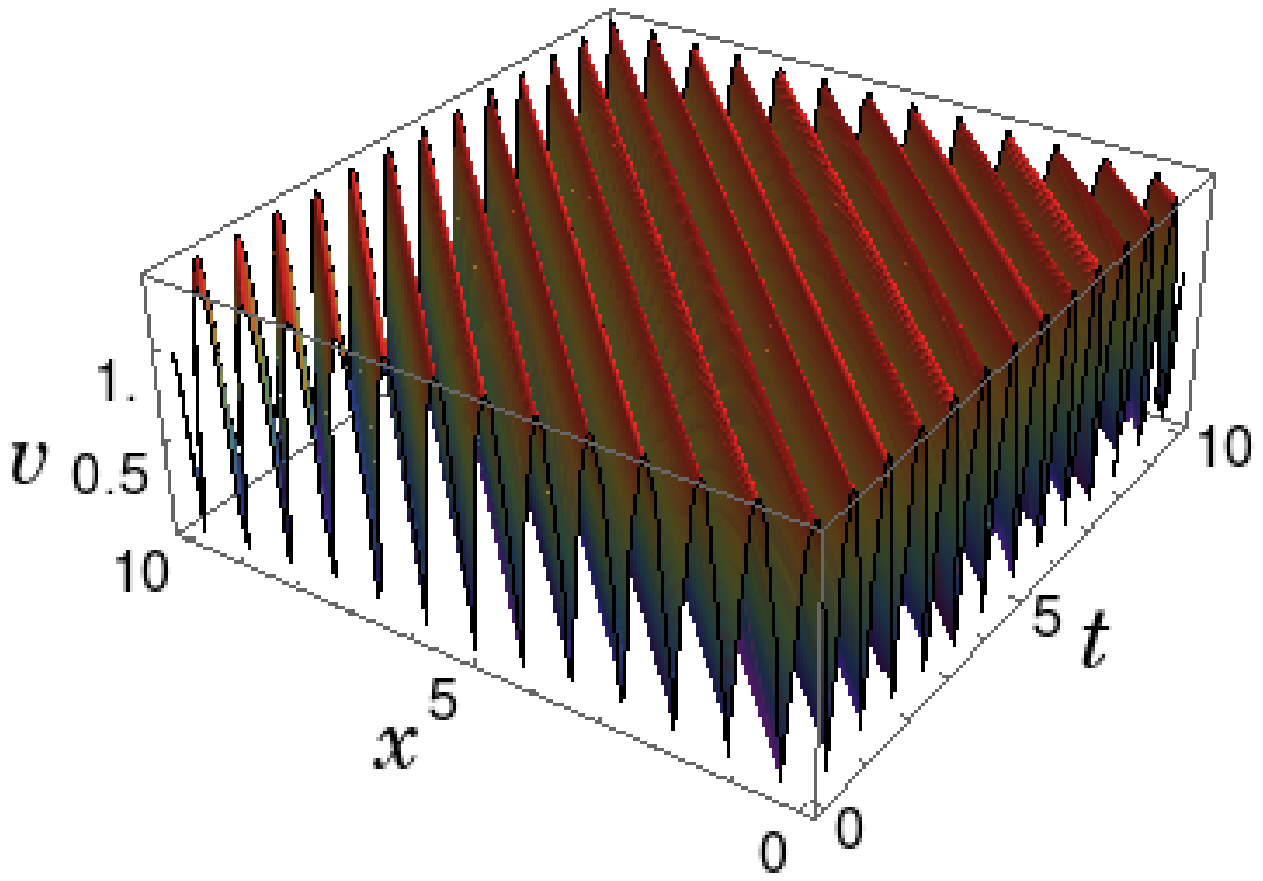}
\caption{Periodic wave solutions of Eq. (\ref{5a}) and the travelling waveform ($y=x-at$) of Eq. (\ref{nlh1}) for $\gamma=2$, $a=1$, $b_1=0.5$ and $b_2=1.5$.}
\label{fg2a}
\end{figure}

For a special case, $k=-1/a^2$, the above system of nonlinear second-order ODEs
(\ref{5}) becomes a set of first-order ODEs
\bes\bny aA'-\g B(A^2+B^2)=0, \\
aB'+\g A(A^2+B^2)=0. \eny\label{5a} \ees
Solving the system of nonlinear ODEs (\ref{5a}), we get the explicit solution
$A(y)=\sqrt{2b_1}\sin((-2b_1 \g /a)y+b_2)$ and
$B(y)=-\sqrt{2b_1}\cos((-2b_1 \g /a)y+b_2)$, where $b_1$ and $b_2$
are arbitrary real constants. Thus a travelling wave solution to the
equations (\ref{nlh1}) is given by
\bes\bny && u(x,t)=\sqrt{2b_1}\sin\(\frac{-2b_1 \g}{a}(x-at)+b_2\), \\
&& v(x,t)=-\sqrt{2b_1}\cos\(\frac{-2b_1 \g}{a}(x-at)+b_2\).\eny\ees
These solutions are not the standard travelling waves due to the dependence of frequency ($2b_1 \g$) and wave number ($2b_1 \g/a$) on the amplitude ($\sqrt{2b_1}$). Also, we have shown the numerically obtained periodic wave solutions in the form of sine and cosine functions for Eq. (\ref{5a}) in Fig. \ref{fg2}. From Fig. \ref{fg2}, we find that the analytical and numerical solutions of Eq. (\ref{5a}) exactly coincide. This ultimately confirms the validity of direct numerical results of other set of ODEs obtained in this work. Such non-standard periodic waves can also be viewed as the propagation of nonlinear waves in bimodal optical fibers. Also in Fig. \ref{fg2a}, we have shown the periodic solution for a different $b_1$ ($b_2$) parameters, which influence the amplitude and frequency (phase-shift). It also contains the original travelling wave solution for Eq. (\ref{nlh1}).\\ %Here we note that the period of oscillations are shorter than that of Eq. (\ref{5}) given in Fig. \ref{fg1}.

\begin{figure}[h]
\centering \includegraphics[width=0.32\linewidth]{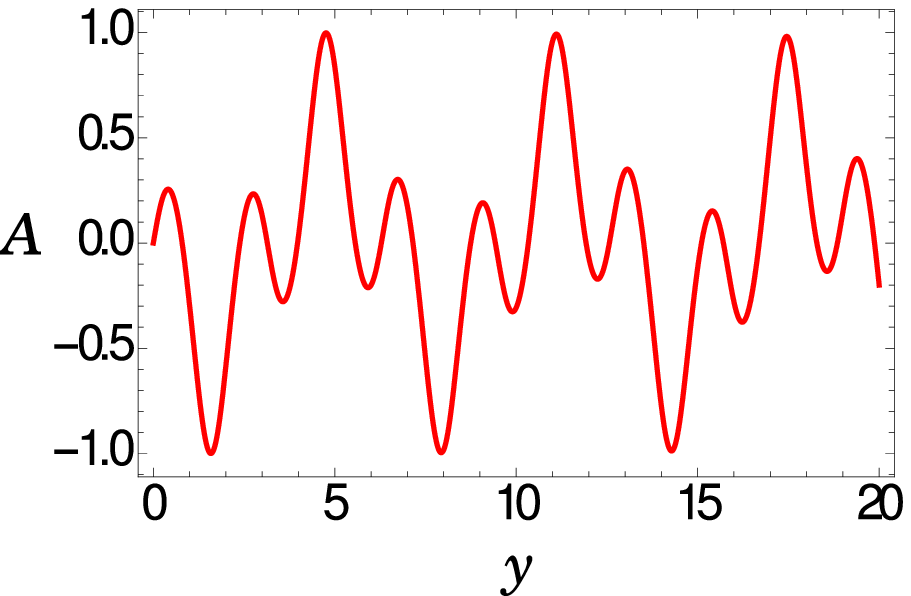}  \includegraphics[width=0.32\linewidth]{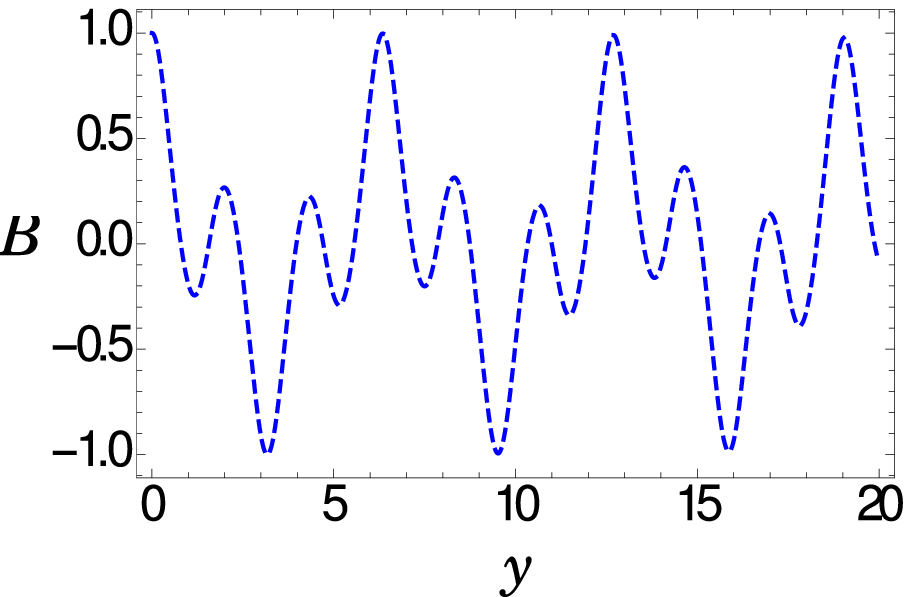}
\caption{Triple peak nonlinear wave trains of Eq. (\ref{5b}) for $k=0.5$ and $\gamma=2$.}
\label{fg3}
\end{figure}

Another subcase of case (i) can be obtained for $a=0$ as $a$ is an arbitrary constant. This choice reduces the subalgebra as $X_1+a X_4 \rightarrow X_1$. Here also we obtain the group-invariant solutions as $u(x,t)=A(y)$ and $v(x,t)=B(y)$ with invariant $y=x$. Consequently the PDEs (\ref{nlh1}) result into the following ODEs:
\bes\bny kA^{''}-B'+\g A(A^2+B^2)=0,  \\
kB^{''}+A'+\g B(A^2+B^2)=0. \eny \label{5b}\ees
The above equation too falls into the category of Painlev\'e integrable case and
the forms of integrals obtained by using the modified
Prelle-Singer method can be written as below, which follows from (\ref{5int}).
\bes\bny	&&I_1=2({A'^2+B'^2})+\frac{\gamma(A^2+B^2)^2}{k},\\
&&I_2={(AB'-A'B)}+\frac{(A^2+B^2)}{2k}.	\eny\ees
The numerical solutions of Eq. (\ref{5b}) are shown in Fig. \ref{fg3} for $k=0.5$ and $\gamma=2$. It shows that system (\ref{5b}) admits triple peak nonlinear wave structures of varying amplitudes. Such type of multipeaked stationary solutions exist in laser-BEC interactions and it has been shown that they can also explain the formation of bound state.
%These multi-peak (multi-hump) structures can be visualized as stationary bright soliton bound states or partially coherent solitons arising in the context of nonlinear optics.
Also, these triple-peak nonlinear wave can be associated with a series of asymmetric optical multipeak solitons on a continuous wave background  \cite{multi-peak}.

\subsection*{{\bf Case (ii)}: $X_2+aX_3$}
\begin{figure}[h]
\centering  \includegraphics[width=0.34\linewidth]{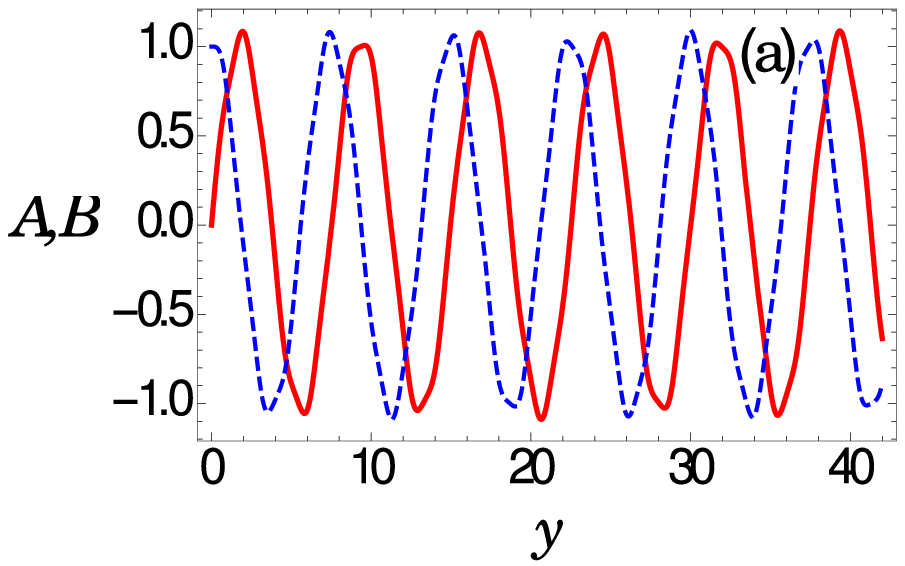}  \includegraphics[width=0.34\linewidth]{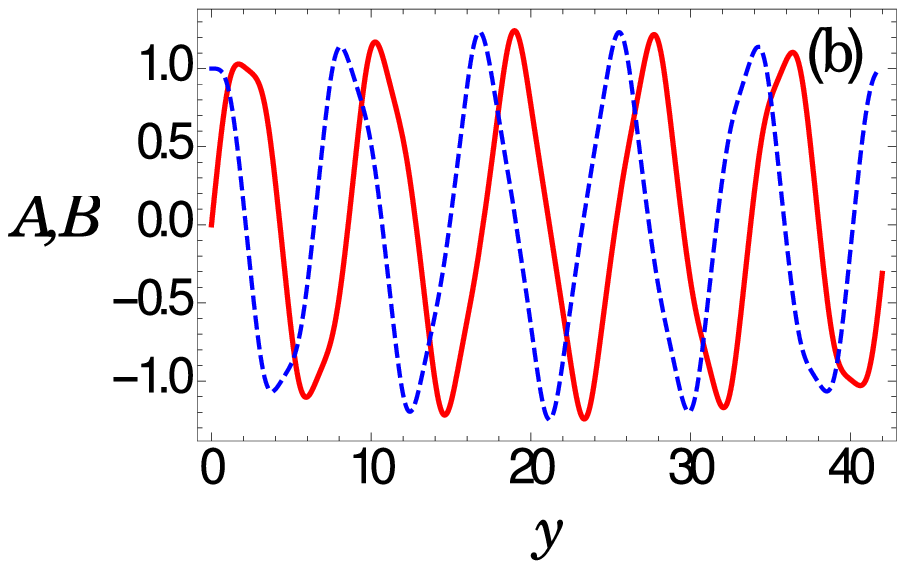}
\caption{Nonlinear wave trains of Eq. (\ref{6}) for $\gamma=2$, $a=1$, (a) $k=0.5$ and (b) $k=1.5$.}
\label{fg4}
\end{figure}

The next subgroup $X_2+aX_3$ in the optimal subalgebras reduces Eq. (\ref{nlh1}) to the following system of nonlinear second-order ODEs
\bes\bny kA^{''}+B'-a^2A+\g A(A^2+B^2)=0,\\
kB^{''}-A'-a^2B+\g B(A^2+B^2)=0. \eny \label{6} \ees
%here `prime' denotes again differentiation with respect to $y$.
The Painlev\'e analysis again shows that Eq. (\ref{6}) is integrable thereby admitting
required number of arbitrary constants for all resonance values.
Again by using the modified Prelle-Singer method we obtain the explicit forms of integrals as below:
\bes\bny	&&I_1=({A'^2+B'^2})+\frac{\gamma}{k}(A^2+B^2)^2-{\frac{2a^2}{k}(A^2+B^2)},\\
&&I_2=(A'B-AB')+\frac{A^2+B^2}{2k}.	\eny \ees
As in the above case, Eq. (\ref{6}) is solved numerically and its solutions are shown in Fig. \ref{fg4}. Here we observe a special set of non-identical chirped pulse trains for $A$ and $B$ which repeats periodically. Such type of chirped pulses find applications in nonlinear optical systems with higher order nonlinearities like quintic and septic effects. Here also one can observe the influence of non-paraxial coefficient $k$, which is directly proportional to the pulse width, that is, the increase in $k$ results in a increase of the pulse width of the nonlinear wave trains (see Fig. \ref{fg4}).

\begin{figure}[h]
\centering \includegraphics[width=0.34\linewidth]{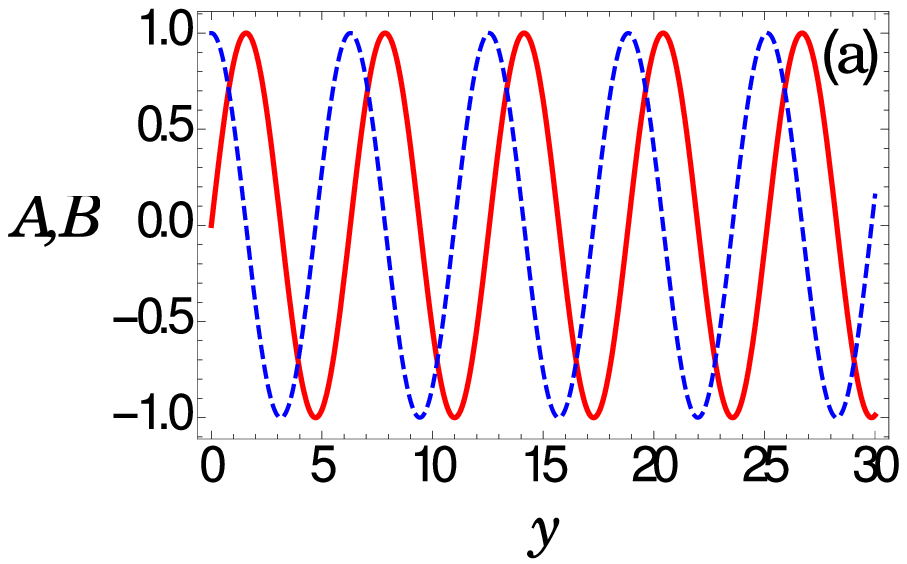}  \includegraphics[width=0.34\linewidth]{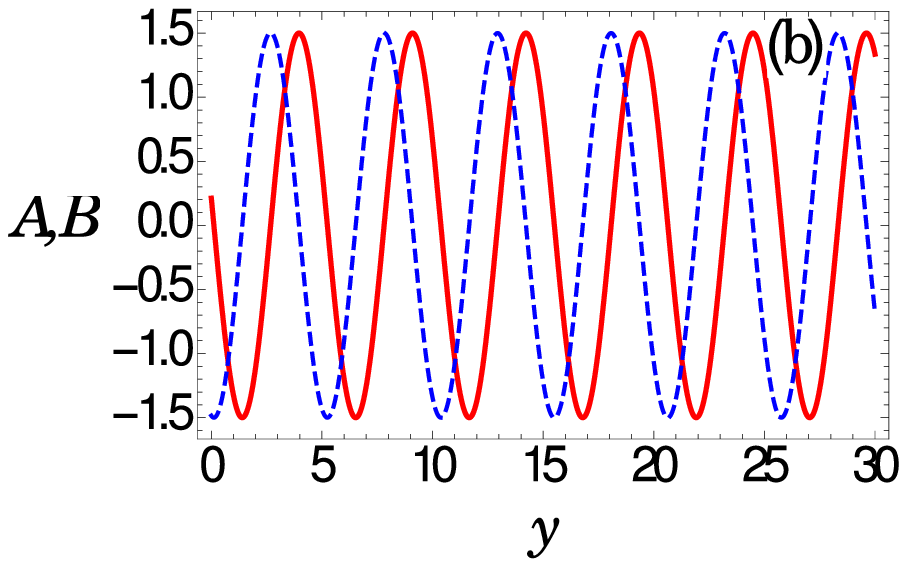}
%\centering \includegraphics[width=0.43\linewidth]{fig5}  \includegraphics[width=0.43\linewidth]{fig5b}
\caption{Periodic sine and cosine wave solutions (\ref{sol6a}) for (a) $b_1=1$ and $b_2=0$ and (b) $b_1=1.5$ and $b_2=3$, with $\gamma=2$.}
\label{fg5}
\end{figure}

One can also obtain a subcase of case (ii) for the choice $a=0$, which results into $X_2+aX_3 \rightarrow X_2$. Here we obtain the group-invariant solutions as $u(x,t)=A(y)$ and $v(x,t)=B(y)$, where the invariant $y=t$, which reduces the PDEs (\ref{nlh1}) into a set of ODEs
\bes\bny A^{''}+\g A(A^2+B^2)=0,  \\
B^{''}+\g B(A^2+B^2)=0. \eny \label{6a}\ees
We have performed the Painlev\'e analysis and it shows that the above equation is P-integrable.
A basic solution for the above equation can be obtained as
\bes\bny
A(t)=b_1 \mbox{sin}(\sqrt{\gamma}b_1 t+b_2),\\
B(t)=b_1 \mbox{cos}(\sqrt{\gamma}b_1 t+b_2),
\eny \label{sol6a}\ees where $b_1$ and $b_2$ are arbitrary constants. We have shown the above periodic wave solutions in Fig. \ref{fg5} for two different choices of $b_1$ and $b_2$. Even though the amplitude of periodic solutions (\ref{sol6a}) are same in $A$ and $B$, they are out of phase to each other. In addition to the above periodic solutions, one can also obtain elliptic function solutions of Eq. (\ref{6a}). %One can note that the numerical solutions are corroborated by the exact analytical solutions, see Fig. \ref{fg5}.

\subsection*{{\bf Case (iii)}: $X_1+aX_2+b X_3$}
From the Table 3, we find that the group-invariant solution resulting for the sub-algebra $X_1+aX_2+b X_3$ gives rise to the following system of nonlinear second-order ODEs to the NLH equations (\ref{nlh1}):
\bes\bny (1+ka^2)A^{''}-a(1+2k b)B'-b (1+k b)A+\g A(A^2+B^2)=0,  \\
(1+ka^2)B^{''}+a(1+2k b)A'-b (1+k b)B+\g B(A^2+B^2)=0. \eny \label{7} \ees
%here `prime' denotes differentiation with respect to $y$.
Here also we find that Eq. (\ref{7}) passes the Painlev\'e test and is integrable for arbitrary values of parameters $k$, $\gamma$, $a$, and $b$.
We have derived the first integrals of Eq. (\ref{7}) by employing the modified Prelle-Singer method as
\bes\bny	&&I_1=\frac{2(1+k b)b}{(1+ka^2)}(A^2+B^2)-\frac{\gamma (A^2+B^2)^2}{(1+ka^2)}-2({A'^2+B'^2}),\\
%&&I_2=2(1+ka^2)(AB'-A'B)+a(1+2k b)(A^2+B^2),
&&I_2=(AB'-A'B)+\frac{a(1+2k b)}{2(1+ka^2)}(A^2+B^2).	
\eny\ees

\begin{figure}[h]
\centering \includegraphics[width=0.33\linewidth]{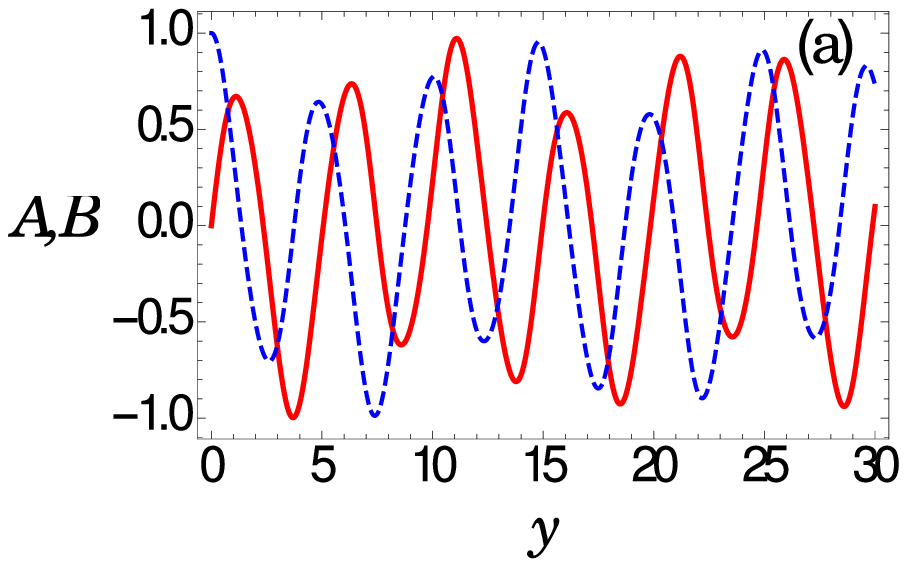} \includegraphics[width=0.33\linewidth]{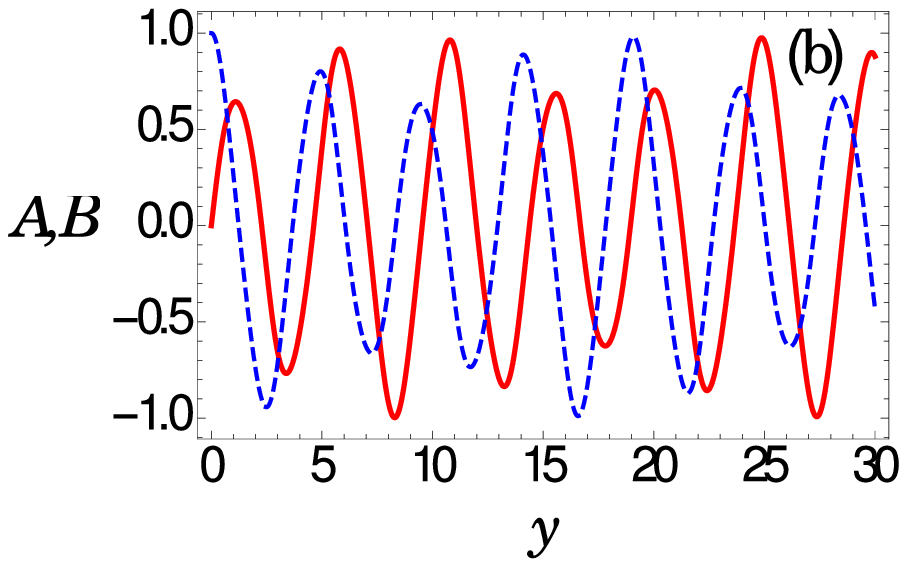} \includegraphics[width=0.33\linewidth]{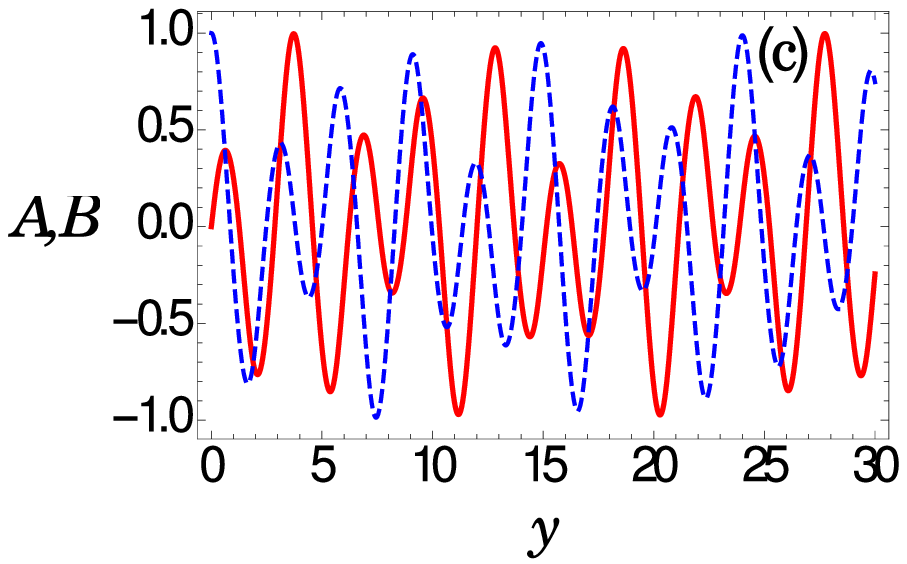}  %\includegraphics[width=0.4\linewidth]{fig6-c}
\caption{Numerically obtained nonlinear wave trains in Eq. (\ref{7}) for (a) $k=0.5$ and $b=1$, (b) $k=0$ and $b=1$, and (c) $k=0$ and $b=-1$, with the other parameters $\gamma=2$ and $a=1$.}
\label{fg6}
\end{figure}

The numerical analysis of Eq. (\ref{7}) displays several types of nonlinear wave trains for different choices of parameters as shown in Fig. \ref{fg6}. The arbitrary system parameters ($a,~b,~k$ and $\g$) can be suitably altered to manipulate the nature of the resulting periodic structures. Here the linear coupling coefficient $b$ changes the periodicity of wave trains in the absence of non-paraxial coefficient ($k=0$) which corresponds to the NLS case.

\begin{figure}[h]
\centering \includegraphics[width=0.3\linewidth]{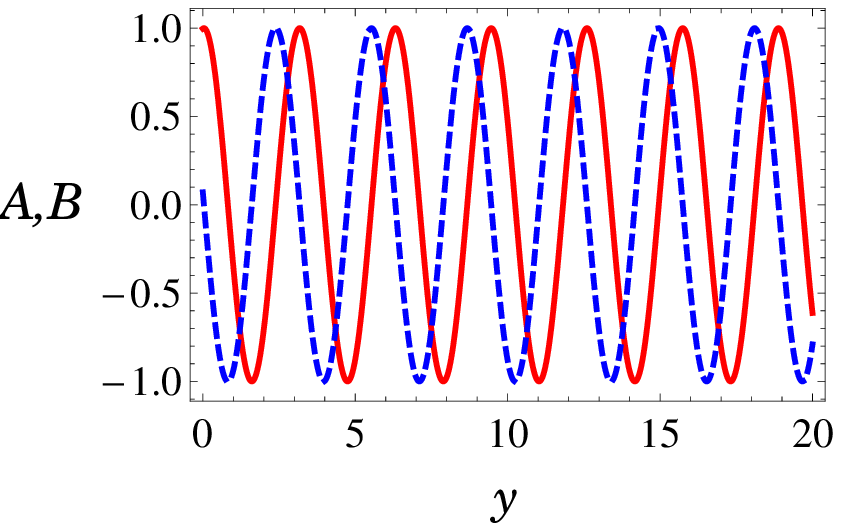}~~~  \includegraphics[width=0.3\linewidth]{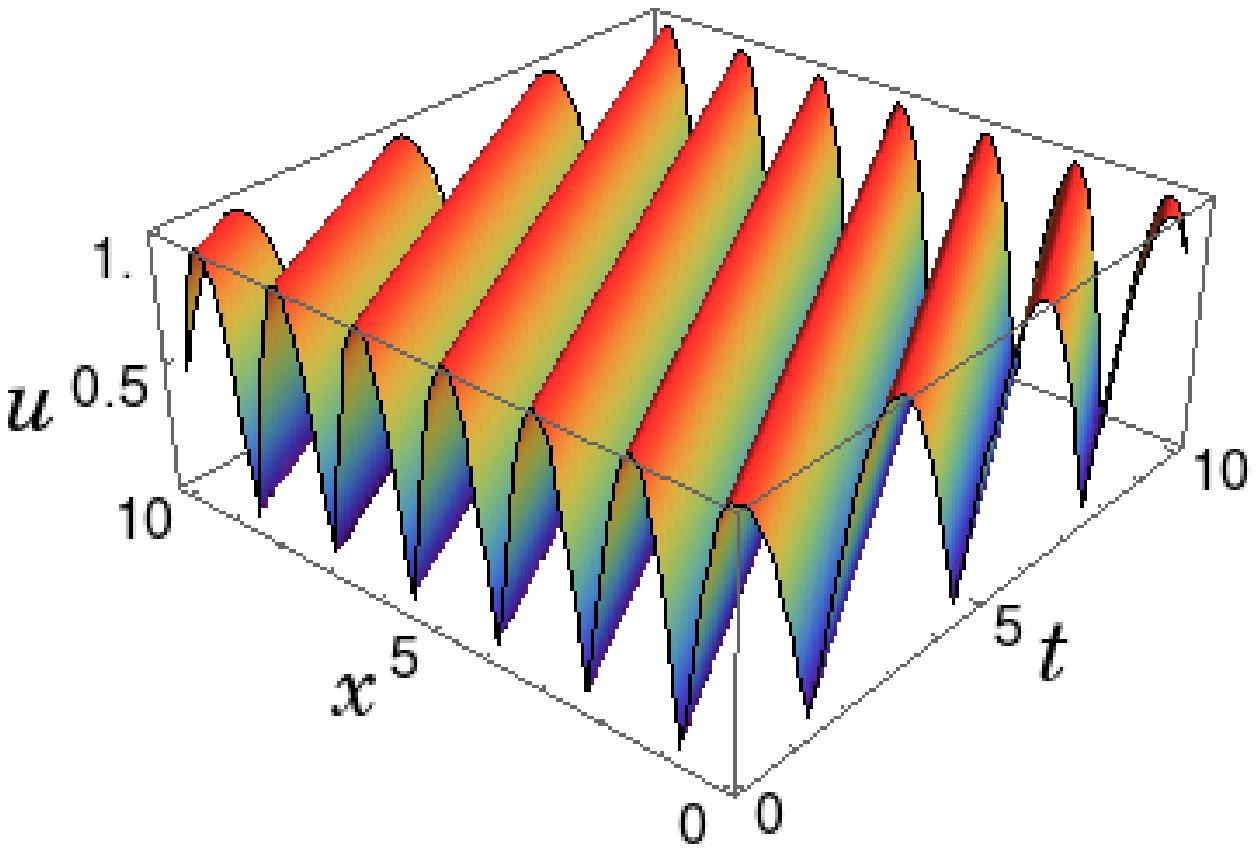} \includegraphics[width=0.3\linewidth]{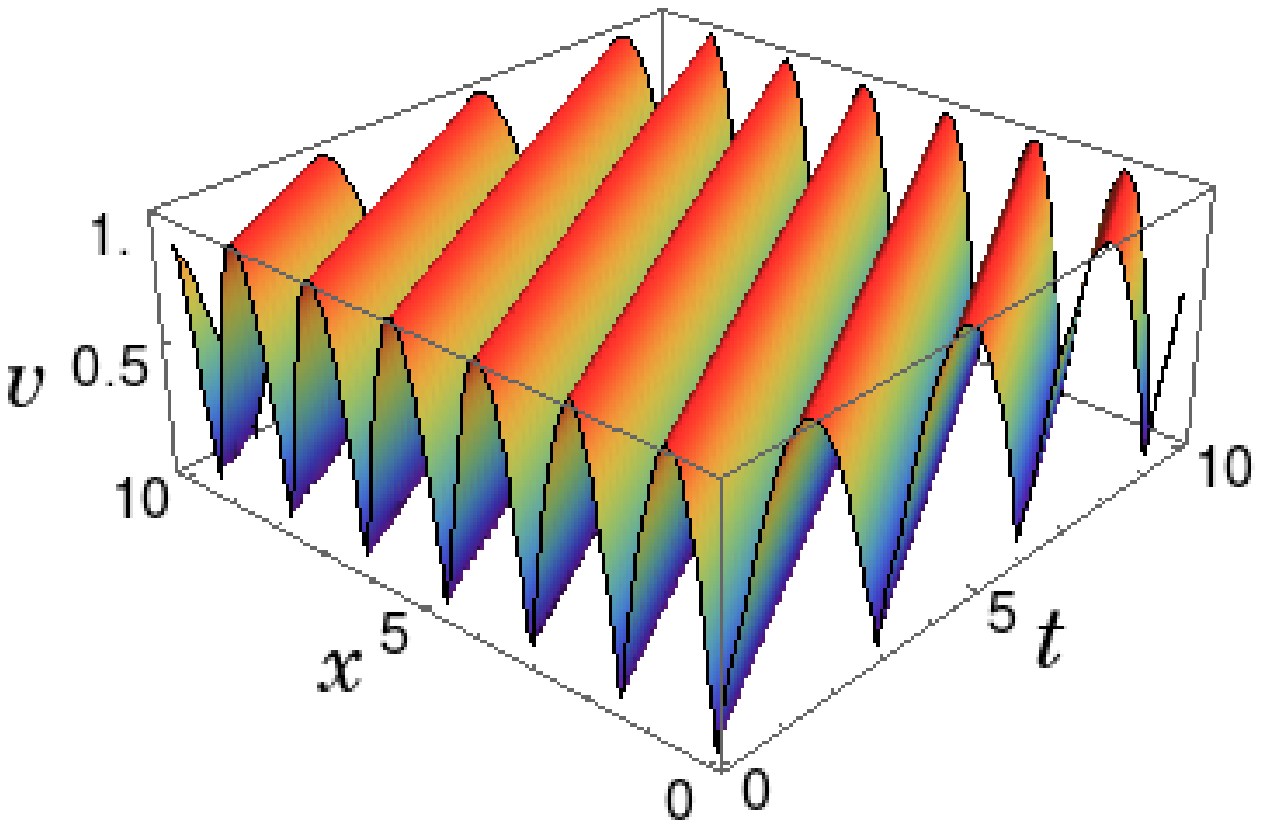}
\caption{Nonlinear periodic wave solution (\ref{sol7a}) of (\ref{7a}) and its travelling waveform ($y=x-at$) of Eq. (\ref{nlh1}) for $\gamma=2$, $a=1$, $b=1$, $b_1=1$ and $b_2=1.5$.}
\label{fg7}
\end{figure}

If $k=-1/a^2$, then the system of nonlinear second-order ODEs
(\ref{7}) becomes
\bes\bny \a B'+\b A-\g A(A^2+B^2)=0,  \\
\a A'-\b B+\g B(A^2+B^2)=0, \eny \label{7a}  \ees
where $\a=a-2 b /a$ and $\b=b-b^2/a^2$.
A solution to the system of nonlinear
first-order ODEs (\ref{7a}) is
$A(y)=\sqrt{b_1}\sin(((\b - \g b_1)/\a)y+b_2)$ and
$B(y)=\sqrt{b_1}\cos(((\b - \g b_1)/\a)y+b_2)$,
where $b_1$ and $b_2$ are arbitrary constants. Thus
the group-invariant solution to the equations (\ref{nlh1}) takes the
form
\bes \bny
u(x,t)=\sqrt{b_1}\cos\(\(\frac{\b - \g b_1}{\a}\)(x-at)+b t+b_2\), \\
v(x,t)=\sqrt{b_1}\sin\(\(\frac{\b - \g b_1}{\a}\)(x-at)+b t+b_2\).\eny \label{sol7a} \ees
We have plotted the above analytical solution of Eq. (\ref{7a}) and the corresponding travelling periodic wave structures of (\ref{nlh1}) in Fig. \ref{fg7}.

\begin{figure}[h]
\centering \includegraphics[width=0.31\linewidth]{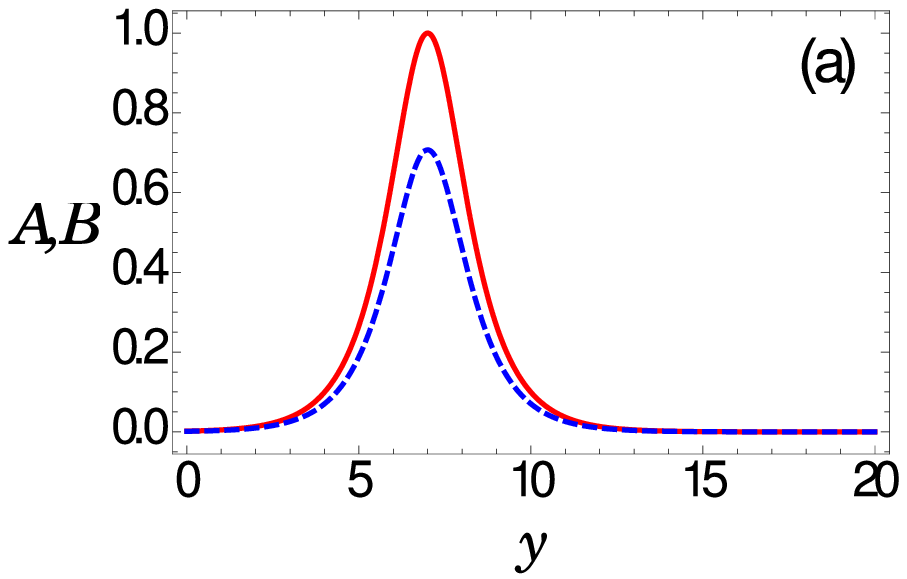}\includegraphics[width=0.31\linewidth]{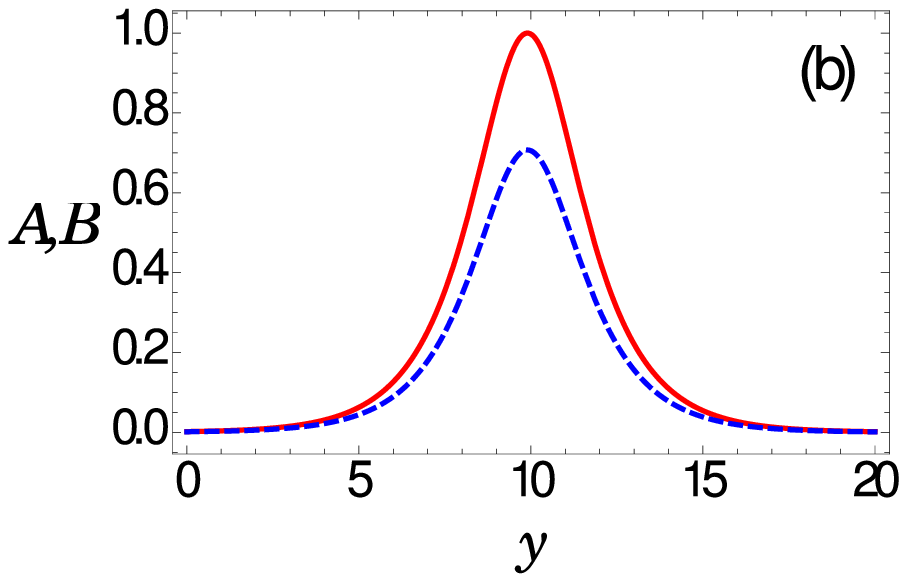} \includegraphics[width=0.31\linewidth]{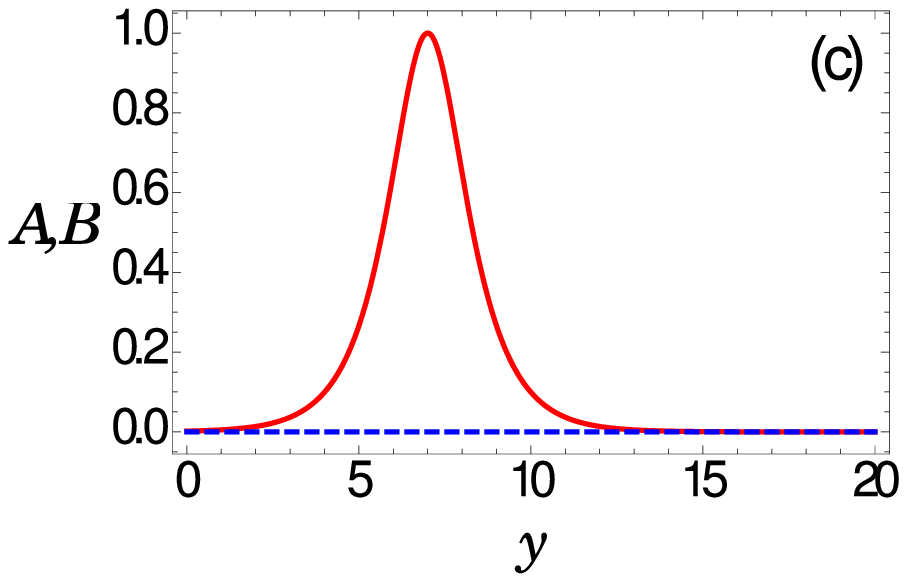}
\caption{Solitary wave solutions, including symbiotic one, of Eq. (\ref{7b}) for (a) $\gamma=2$, and $a=\pm 1$, (b) $\gamma=2$, and $a=0$, and (c) $\gamma=1$, and $a= 1$, along with $b=1$, $k_2=-7$.}
\label{nlh-sol}
\end{figure}

One can also get another special case of Eq. (\ref{7}) for $k=\frac{-1}{2b}$, where the resultant set of ODEs become
\bes\bny
(1+ka^2)A^{''}-b (1+k b)A+\g A(A^2+B^2)=0,  \\
(1+ka^2)B^{''}-b (1+k b)B+\g B(A^2+B^2)=0.
\eny \label{7b} \ees
%The above Eq. (\ref{7b}) exhibits Jaccobian elliptic function solutions
%Also, Eq. (\ref{7b})
The above Eq. (\ref{7b}) admits a special bright solitary wave solution which can be written in the following form.
\bes\bea
&&A=a_1~ \mbox{sech}(k_1 y+k_2),\\
&&B=a_2~ \mbox{sech}(k_1 y+k_2),
\eea \label{sol7b}\ees
with the constraints $k_1^2={\frac{b(1+kb)}{1+ka^2}}$ and $a_2^2={\frac{2b(1+kb)}{\g}-a_1^2}$, where we take $k_1^2> 0$. In Fig. \ref{nlh-sol} we have shown the above bright solitary wave solution for a focussing type nonlinearity ($\g>0$). Note that the width as well as the central position of the solitary waves alter as we change the parameter `$a$'  without affecting their amplitudes. Interestingly, for particular a choice of parameters, namely ${{2b(1+kb)}={\g}a_1^2}$, the solitary wave appears only in the first component ($A$) while it vanishes in the other ($B$) component, see Fig. \ref{nlh-sol}c. We refer to these structures as symbiotic solitary waves. Here we expect that Eq. (\ref{7b}) can also support dark solitary wave for defocussing type nonlinearity ($\g<0$). It would be nice to have rigorous study on (\ref{7b}) in future dealing with various solutions.

The above solutions (\ref{sol7b}) display interesting travelling solitary wave structures with $y=x-at$ in the NLH system (\ref{nlh1}) which we have shown below. In fact we observe beating solitary wave propagation, where the amplitude of solitary wave oscillates periodically.
\begin{figure}[h]
\centering \includegraphics[width=0.32\linewidth]{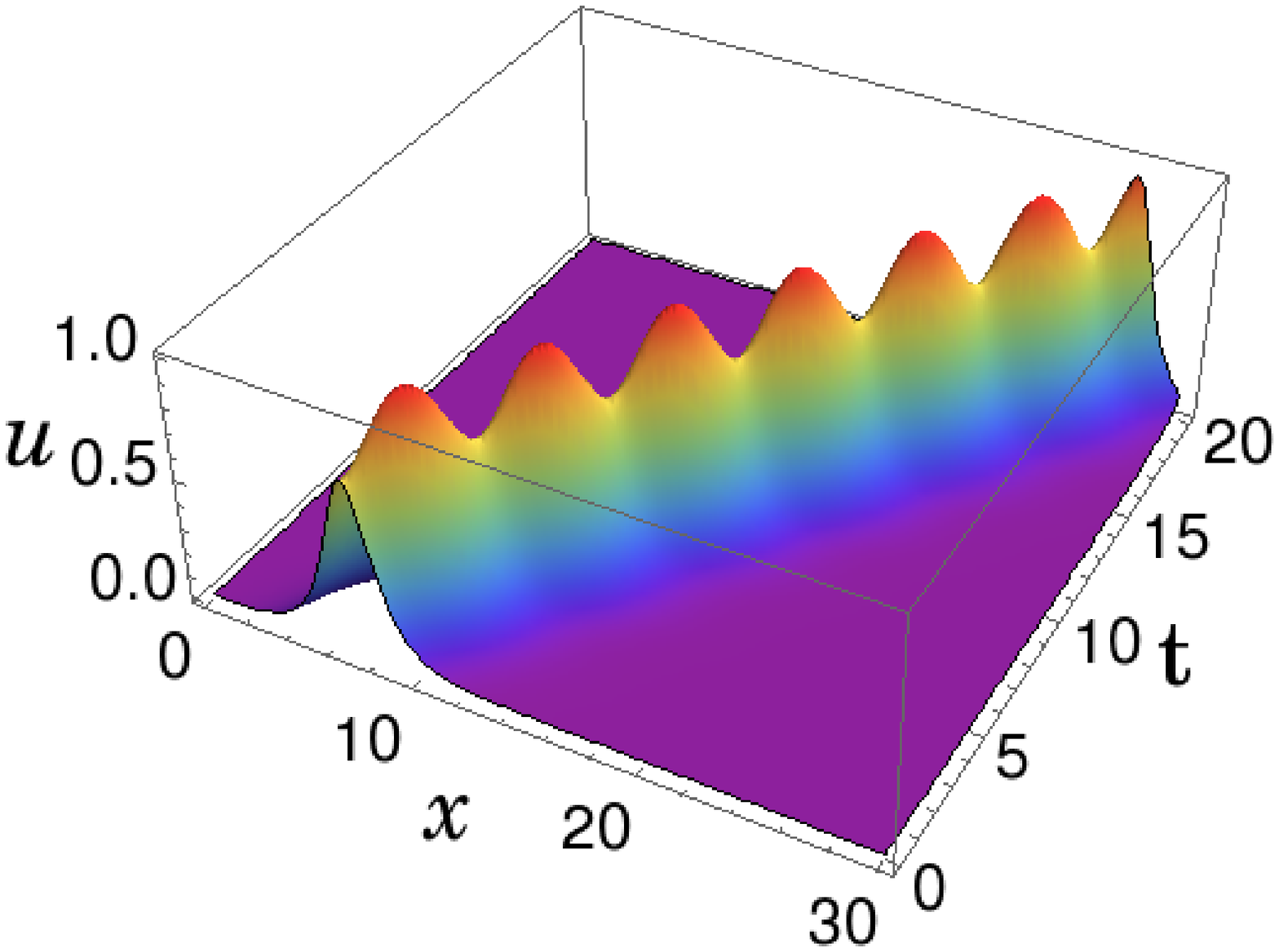} ~~~\includegraphics[width=0.32\linewidth]{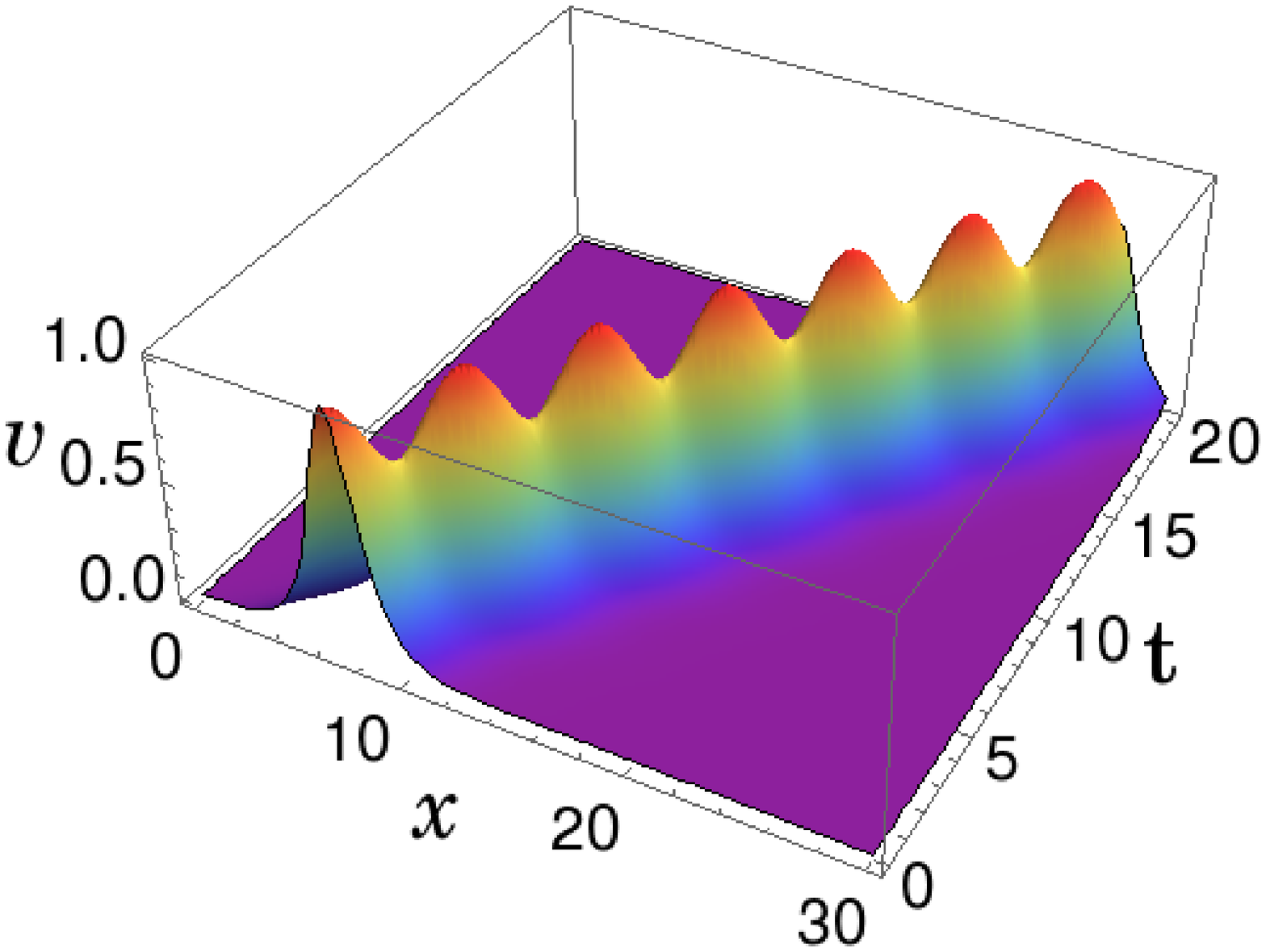}
\caption{Travelling solitary wave solutions of Eq. (\ref{nlh1}) for the choice in Fig. \ref{nlh-sol}a with $y=x-at$.}
\label{nlh-sol2}
\end{figure}
\begin{figure}[h]
\centering \includegraphics[width=0.32\linewidth]{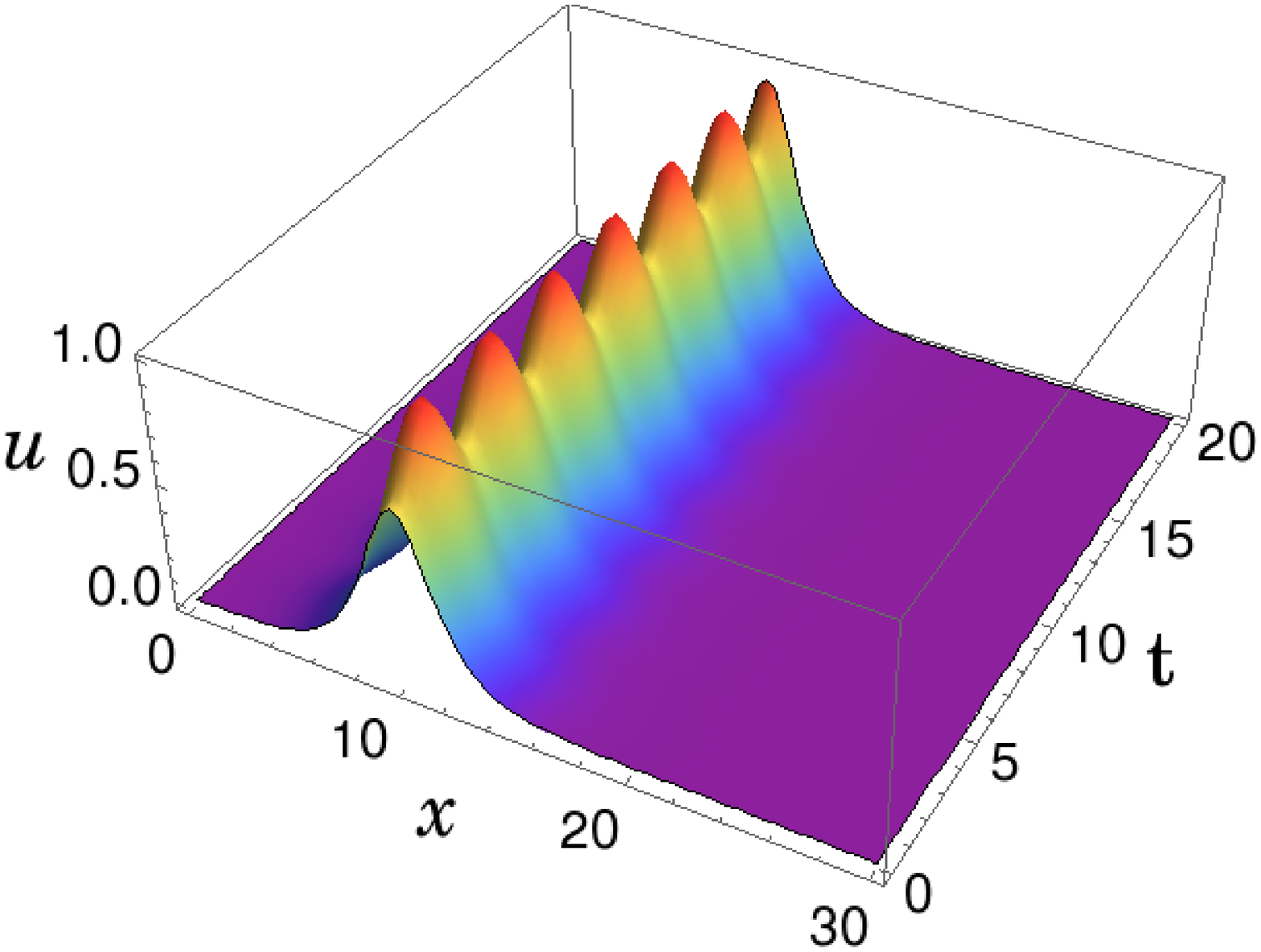} \includegraphics[width=0.32\linewidth]{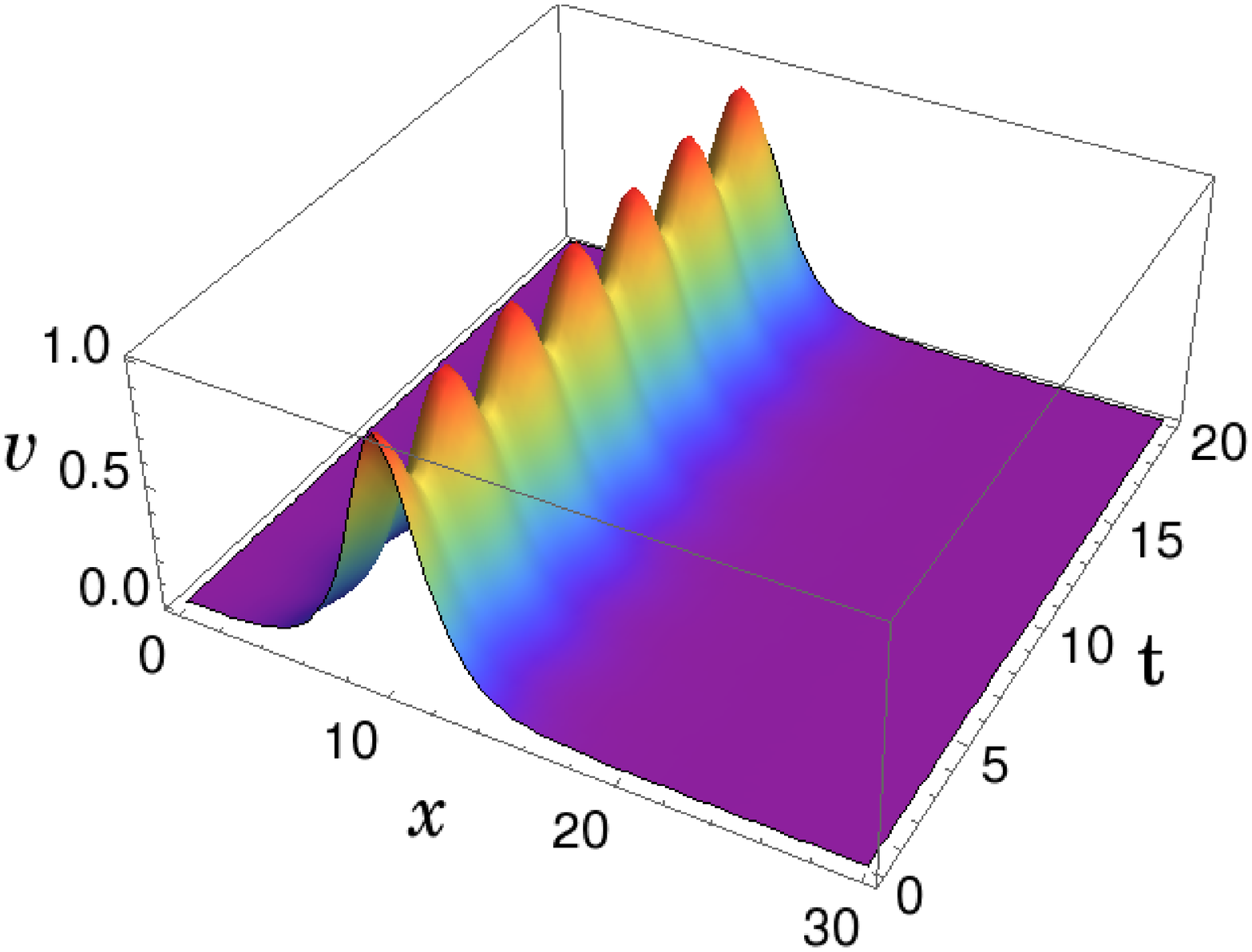}%\includegraphics[width=0.32\linewidth]{fig8-h}
\caption{Stationary solitary wave solutions of Eq. (\ref{nlh1}) for the choice in Fig. \ref{nlh-sol}b with $y=x-at$.}
\label{nlh-sol3}
\end{figure}

%\end{document}
\subsection*{{\bf Case (iv)}: $X_4$}
\begin{figure}[h]
\centering \includegraphics[width=0.32\linewidth]{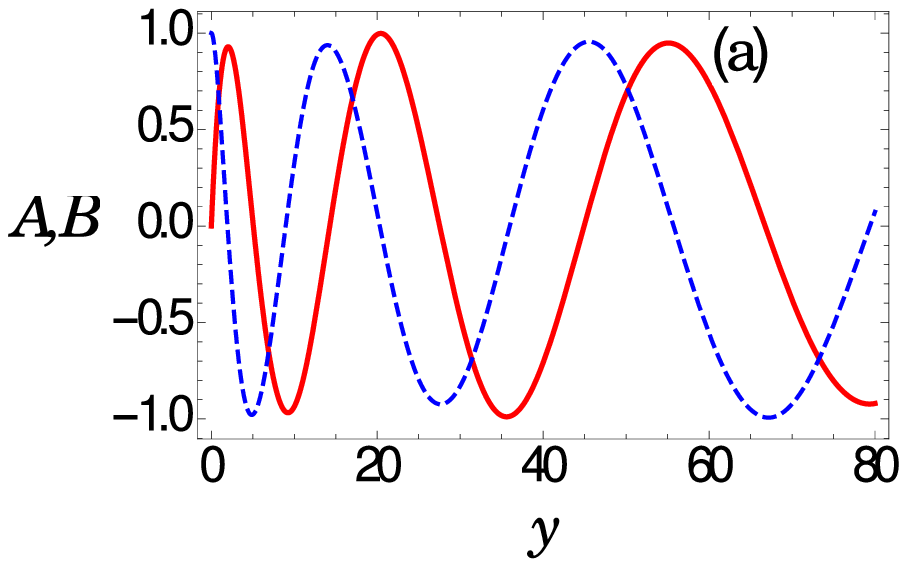}  \includegraphics[width=0.32\linewidth]{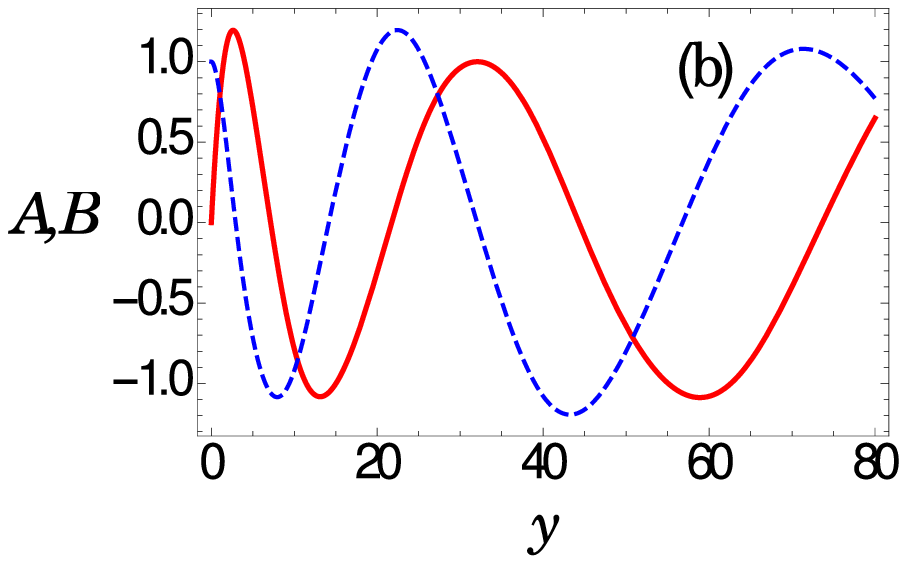}
\caption{Numerical solutions of Eq. (\ref{8}) for (a) $k=0.5$ and (b) $k=1.0$ with $\gamma=2$.}% An exponential chirp waveform; a sinusoidal wave that decreases in frequency exponentially over time}
\label{fg8}
\end{figure}

Finally, for the symmetry generator $X_4$ and its group-invariant solution, the system of PDEs (\ref{nlh1}) can be reduced into a system of nonlinear second-order ODEs as
\bes\bny yA^{''}+\frac{1}{2}A'+\frac{1}{16k^2}A+\frac{\g}{4k} A(A^2+B^2)=0,  \\
yB^{''}+\frac{1}{2}B'+\frac{1}{16k^2}B+\frac{\g}{4k} B(A^2+B^2)=0.
 \eny \label{8} \ees
%here `prime' denotes differentiation with respect to $y$.
Our analysis shows that the above set of ODEs (\ref{8}) is also Painlev\'e integrable.
By applying the modified Prelle-Singer method \cite{ePS},
we obtain the first integrals of the above equations (\ref{8}) as
\bes\bny	&&I_1=\frac{(A'^2+B'^2)y}{2}+\frac{\gamma(A^2+B^2)^2}{16k}+\frac{(A^2+B^2)}{32k^2},\\
&&I_2={y^{1/2}}(A'B-AB').	\eny\ees
The direct numerical analysis of Eq. (\ref{8}) is carried out and the results are given in Fig. \ref{fg8}. It shows that due to the explicit appearance of `$y$' in the model equation (\ref{8}), the nonlinear wave train expands almost with constant amplitude as $y$ increases. %period of oscillations is continuously increasing.

In spite of all the above similarity reduced ODEs being integrable, one can show that the original NLH system is non-integrable. More details will be presented elsewhere.
%From the above analysis, we wish to mention that the NLH equation (\ref{0}) is non-integrable in general.
%However, we have shown that the several forms of its symmetry reductions are all integrable by performing the Painlev\'e analysis and also obtained necessary constants of motion.

\section{Lie symmetry analysis of the nonlinear Schr\"odinger equation (\ref{nls})}
As mentioned in the introduction, in this section we briefly outline the results of  symmetry reductions of the NLS equation (\ref{nls}) which arises as a special case of the NLH equation (\ref{0}), for comparison purpose. We wish to note that the symmetry analysis of the NLS system has already been carried out in detail in refs. \cite{winter,nls-res}. %and subsequently by several variants of NLS system have been studied  \cite{AGP,lie6,MS,lie-nls-other}. 
By rewriting Eq. (\ref{nls}) into real and imaginary parts, we get a set of PDEs as given in (\ref{nlh1}) with $k=0$.
\bes\bny u_t+v_{xx}+\gamma\,v(u^2+v^2)=0, \\
v_t-u_{xx}-\gamma\,u(u^2+v^2)=0. \eny \label{nls1} \ees
By applying the infinitesimal symmetry transformations (\ref{trans}) and recursively solving the equations resulting from (\ref{nlh1}), 
%Proceeding as in the case of NLH system here we obtain the forms of infinitesimal functions as below:
\bes \bny
&&\tau=c_1+2c_5 t, \qquad \qquad\qquad
\xi=c_2+2c_4 t+c_5 x , \\
&&\eta^1=-(c_3+c_4 x)v-c_5 u ,\qquad
\eta^2=(c_3+c_4 x)u-c_5 v.
\eny \label{co-nls}\ees
Then Eq. (\ref{co-nls}) leads to the following symmetry generators for the NLS equation \eqref{nls}
\bes\bny &&X_1=\p_t,~~~X_2=\p_x,~~~X_3=-v\,\p_u+u\p_v,\\
&&X_4 =2t\,\p_x-xv\,\p_u+xu\,\p_v,  \\
&&X_5 =2t\,\p_t+x\,\p_x-u\,\p_u-v\,\p_v.
\eny \label{4nls} \ees
%\st{which have been well studied in the literature} \cite{PK-ML,nls-res}. \st{Here, we only present the minimum details to compare with NLH case.}
From the above generators we can note that the NLS equation (\ref{nls}) is invariant under translation
in time ($X_1$), translation in space ($X_2$), transformation of the phases ($X_3$), Galilean boost ($X_4$), and scaling transformation ($X_5$).
From the optimal system of one-dimensional subalgebras, we find that the following
combinations of generators give group invariant solutions for the NLS equation (\ref{nls}):
(i) $X_5+aX_3$, (ii) $X_1+c X_2$, (iii) $X_2+c X_3$,
(iv) $X_1+c X_2+b X_3$ (v) $X_4+c X_1$ and (vi) $X_3$, where
$a$, $b$ and $c$ are arbitrary real constants.\\

%\end{document}
\noindent{\bf Case (i)}: The symmetry generator $X_5+a X_3$ gives the
group-invariant solution of the form
$u(x,t)=t^{-1/2}[A(y)\cos(a/2\, \ln t)+B(y)\sin(a/2\, \ln t)]$ and
$v(x,t)=t^{-1/2}[A(y)\sin(a/2\, \ln t)-B(y)\cos(a/2\, \ln t)]$
with $y=xt^{-1/2}$.
This reduces the system of PDEs (\ref{nls1}) to the following system
of nonlinear second-order ODEs
\bes\bny
2A^{''}-aA-yB'-B+2\g A(A^2+B^2)=0, \\
2B^{''}-aB+yA'+A+2\g B(A^2+B^2)=0. \eny \label{5nls} \ees
%here `prime' denotes differentiation with respect to $y$.
The above system of ODEs is found to be Painlev\'e integrable
and sufficient number of independent integrals can be found as in the case of Eq. (\ref{5}) to show the integrability.
The direct numerical analysis of the above coupled ODEs is shown below in Fig. \ref{fgnls}. Here the explicit appearance of $y$ in the first derivative term leads to a compression of wave train as $y$ increases Fig. \ref{fgnls}(a)-(b). Meanwhile, the amplitude reaches a steady state. This resembles the behaviour of an attractor. %Here the pulse compression takes place due to the linear cross coupling term in Eq. (\ref{5nls}), namely $B$ in (\ref{5nls}a) and $A$ in (\ref{5nls}b).
In the absence of linear coupling terms, one can obtain a periodic waves with diminishing amplitude as $y$ increases. We have shown such nonlinear wave trains in Fig. \ref{fgnls}(c)-(d).  \\
\begin{figure}[h]
\centering \includegraphics[width=0.32\linewidth]{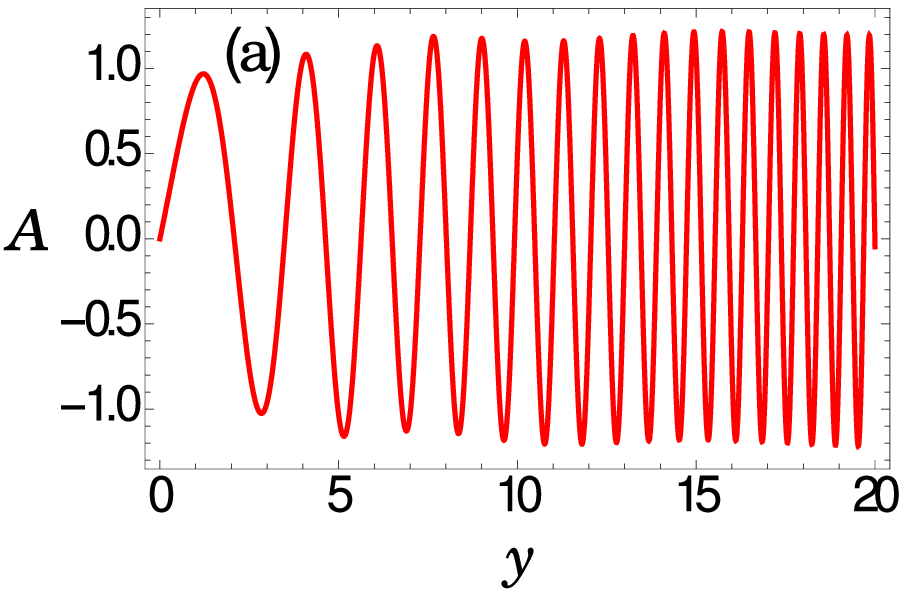}  \includegraphics[width=0.32\linewidth]{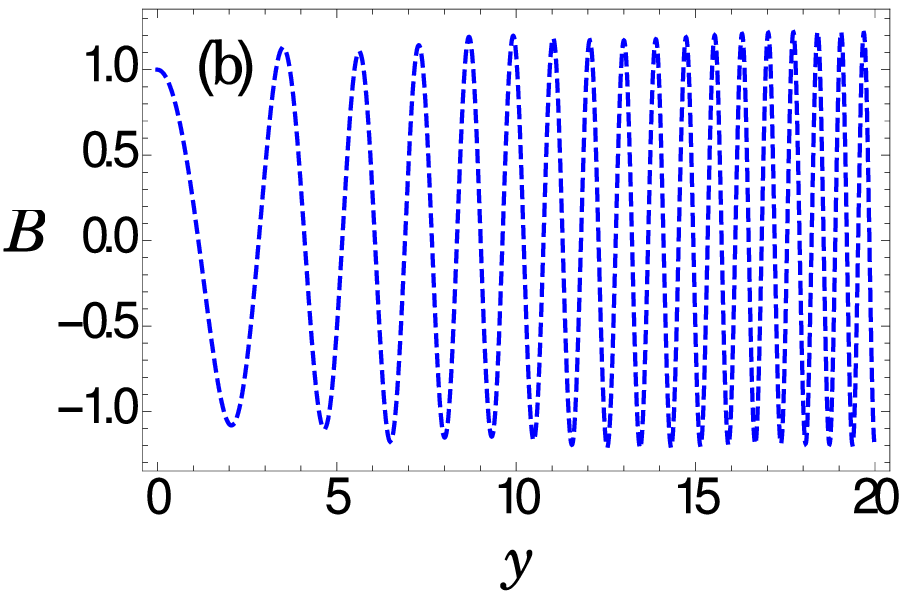}\\
\centering \includegraphics[width=0.32\linewidth]{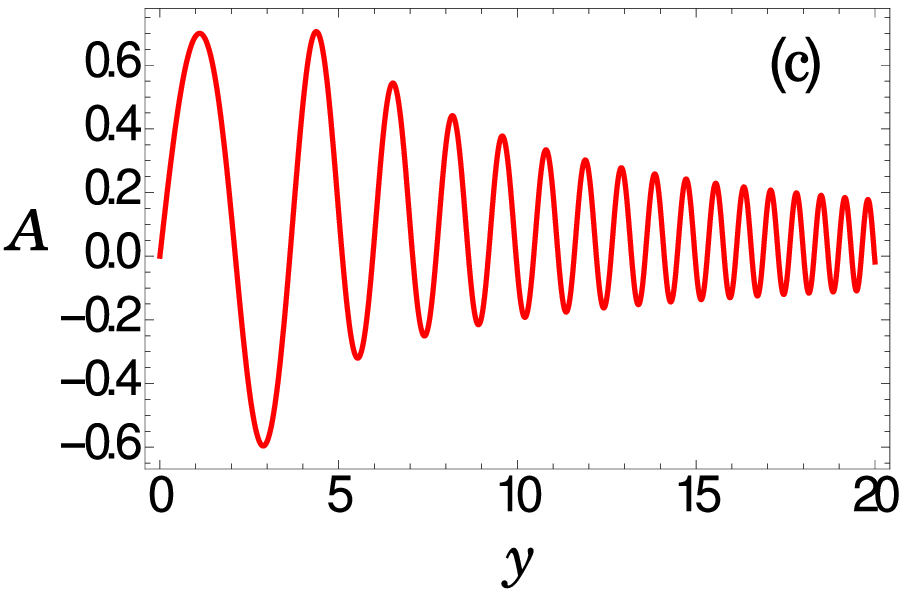}  \includegraphics[width=0.32\linewidth]{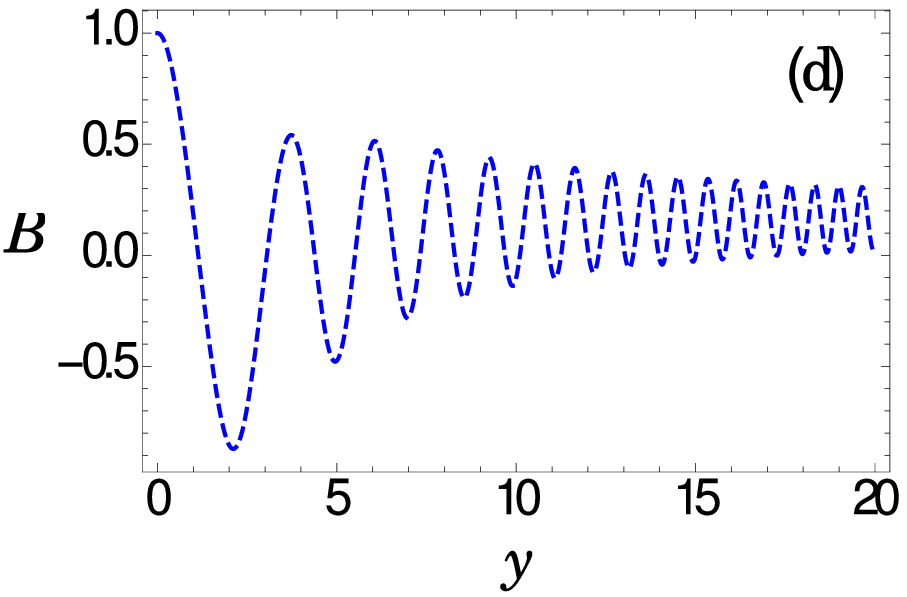}
\caption{Numerical nonlinear wave trains in Eq. (\ref{5nls}) $\gamma=2$. The upper and lower panels show the wave trains in the presence ($a=1$) and absence ($a=0$)  of linear coupling terms. }% An exponential chirp waveform; a sinusoidal wave that decreases in frequency exponentially over time}
\label{fgnls}
\end{figure}

\noindent{\bf Case (ii)}: For the sub-algebra with generators $X_1+c X_2$,
one can obtain the group-invariant solution
$u=A(y)$ and $v=B(y)$ with the invariant $y=x-c t$.
From this, the system of PDEs (\ref{nls1}) can be reduced to the system of
nonlinear second-order ODEs,
\bes\bny
A^{''}+c B'+\g A(A^2+B^2)=0, \\
B^{''}-c A'+\g B(A^2+B^2)=0. \eny \label{6nls}\ees
%here `prime' denotes differentiation with respect to $y$.
Note that the above equation (\ref{6nls}) also passes the Painlev\'e test and becomes integrable. We wish to mention that the above equation (\ref{6nls}) is similar to Eq. (\ref{5b}) and we can obtain multi-peak nonlinear wave structures as shown in Fig. \ref{fg3}.

If $c =0$, the group-invariant solution corresponding to $X_1$
of the system of PDEs (\ref{nls1}) is $u=A(y),~v=B(y)$, where $y=x$,
and $A(y)$ and $B(y)$ satisfy the following system of nonlinear
second-order ODEs,
\bes\bny
A^{''}+\g A(A^2+B^2)=0, \\
B^{''}+\g B(A^2+B^2)=0.\eny  \label{7nls} \ees
Here we find that the above system of ODEs (\ref{7nls}) is the same as the integrable equation (\ref{6a}) and it admits periodic as well as elliptic function solutions which have already been addressed. \\

\noindent{\bf Case (iii)}: In this case, for $X_2+c X_3$, we obtain
the group-invariant solutions
$u=-A(y)\sin c x+B(y)\cos c x$ and $v=A(y)\cos c x+B(y)\sin c x$
with the invariant $y=t$. From this and Eq. (\ref{nls1}), we get the
following set of coupled nonlinear first-order ODEs,
\bes\bny A^{'}+c^2 B-\g B(A^2+B^2)=0, \\
B^{'}-c^2 A+\g A(A^2+B^2)=0. \eny \label{8nls} \ees
%where `prime' means differentiation with respect to $y$.
Equation (\ref{8nls}) is equivalent to Eq. (\ref{7a}) and exhibits the standard periodic wave solutions as shown in Fig. \ref{fg7}.
One can solve the above system of ODEs (\ref{8nls}) and
obtain the following form of group-invariant solutions for (\ref{nls1})
\bes\bny
u(x,t)=\sqrt{b_1}\cos((\g b_1-c^2)t+c x+b_2), \\
v(x,t)=\sqrt{b_1}\sin((\g b_1-c^2)t+c x+b_2), \eny \ees
where $b_1$ and $b_2$ are constants.

If $c =0$, the group-invariant solution corresponding to $X_2$
of the system of PDEs (\ref{nls1}) is $u=A(y),~v=B(y)$, where $y=t$,
and $A(y), B(y)$ satisfy the following system of nonlinear
first-order ODEs,
\bes\bny
A^{'}+\g B(A^2+B^2)=0, \\
B^{'}-\g A(A^2+B^2)=0. \eny \label{9nls} \ees
The above system of ODEs (\ref{9nls}) is similar to Eq. (\ref{5a}).
The solution of (\ref{9nls}) gives rise to the following group-invariant solution for the system of PDEs (\ref{nls1}), which is independent of $x$,
\bes \bny
u(x,t)=\sqrt{b_1}\cos(\g b_1 t+b_2), \\
v(x,t)=\sqrt{b_1}\sin(\g b_1 t+b_2),\eny\ees
where $b_1$ and $b_2$ are constants.\\

\noindent{\bf Case (iv)}: For $X_1+b X_3+c X_2$, we obtain the invariant
$y=x-c t$ and the corresponding invariant solution
$u=-A(y)\sin b t+B(y)\cos b t$ and $v=A(y)\cos b t+B(y)\sin b t$.
This invariant solution reduces the system of PDEs (\ref{nls1}) to the system
of coupled nonlinear second-order ODEs,
\bes\bny A^{''}-b A-c B'+\g A(A^2+B^2)=0, \\
B^{''}-b B+c A'+\g B(A^2+B^2)=0.\eny  \label{10nls}\ees
%here `prime' denotes differentiation with respect to $\g$.
Equation (\ref{10nls}) is equivalent to Eq. (\ref{7}) for $k=0$ and is also Painlev\'e integrable.
Here one can obtain different kinds of nonlinear wave trains for various choices of the arbitrary constants $b$ and $c$.

\begin{figure}[h]
\centering \includegraphics[width=0.33\linewidth]{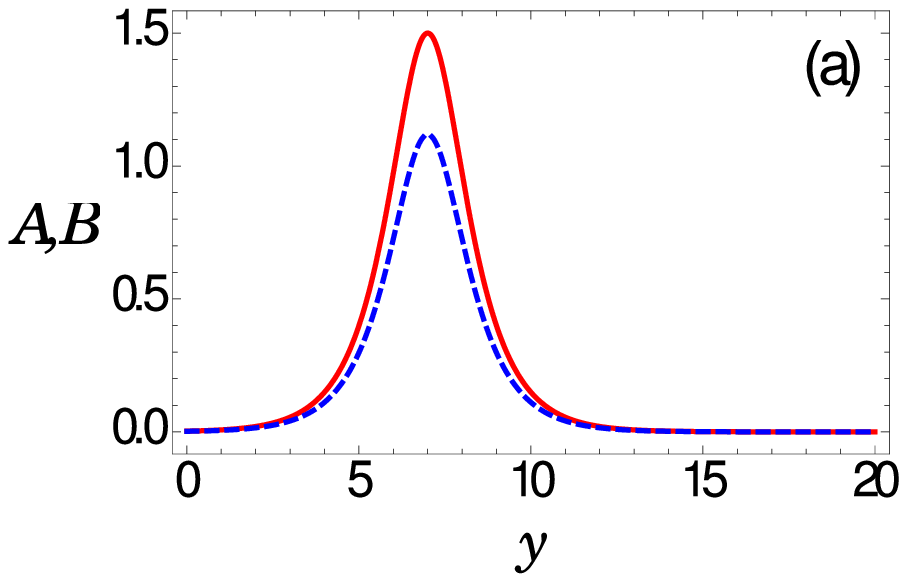} ~~\includegraphics[width=0.33\linewidth]{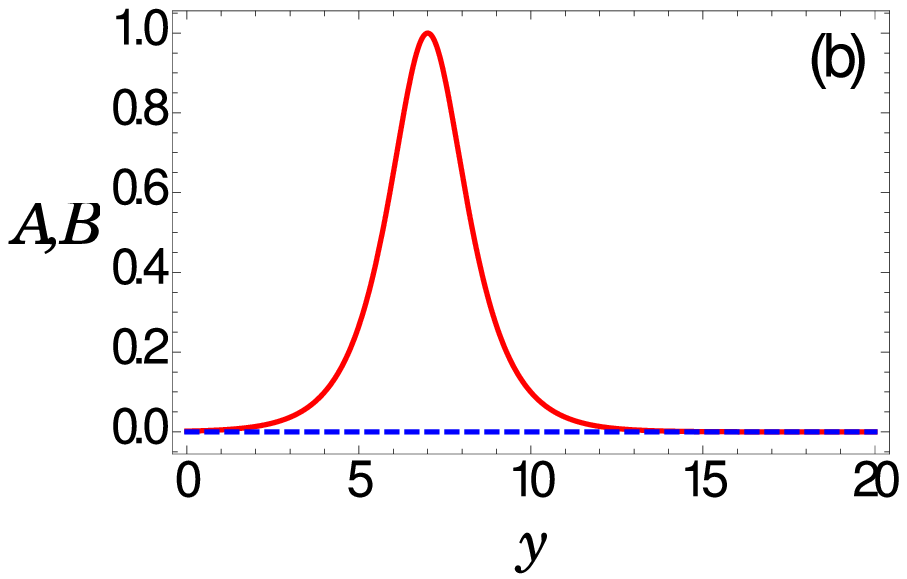}\\
\centering \includegraphics[width=0.33\linewidth]{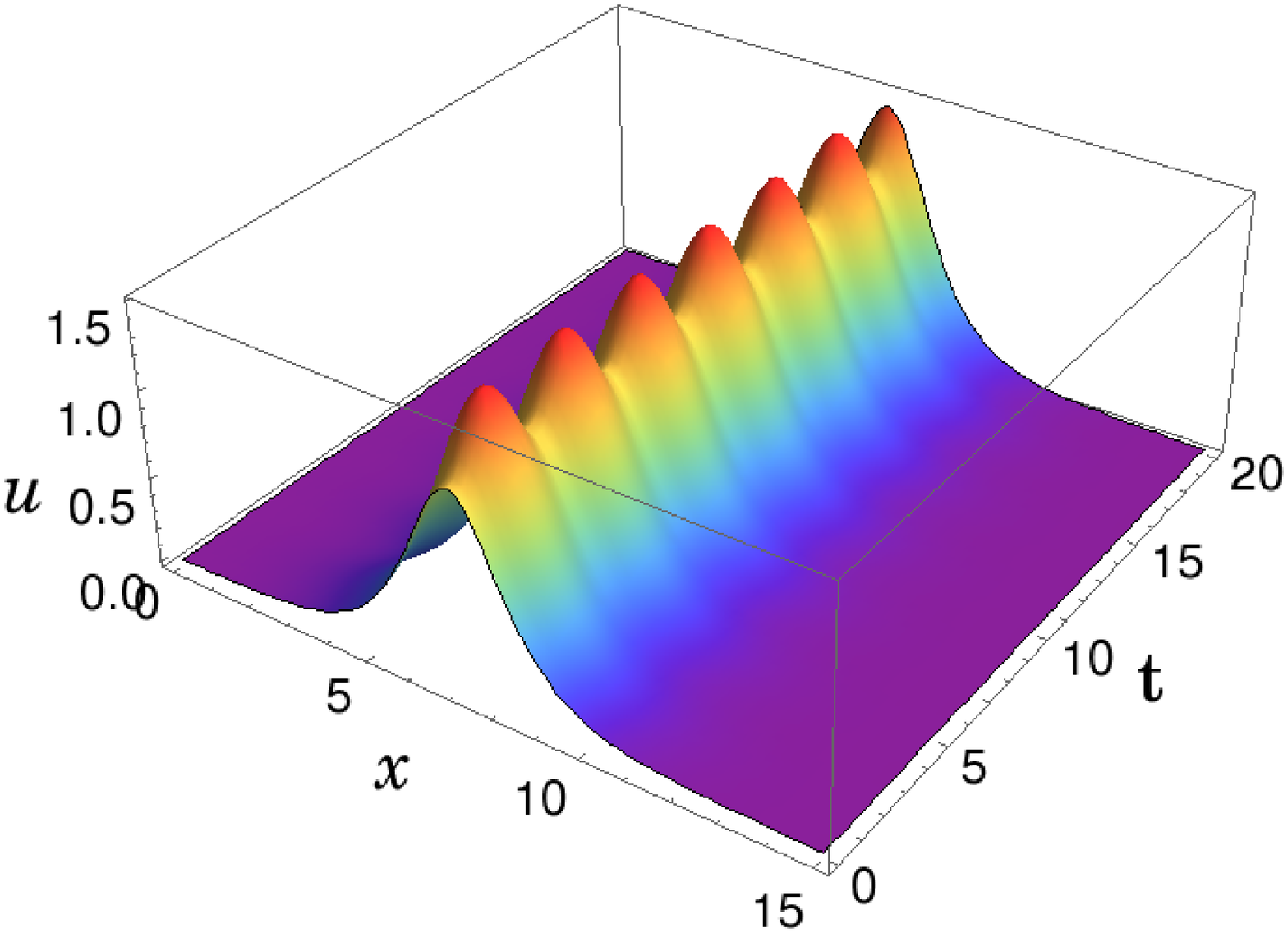} \includegraphics[width=0.33\linewidth]{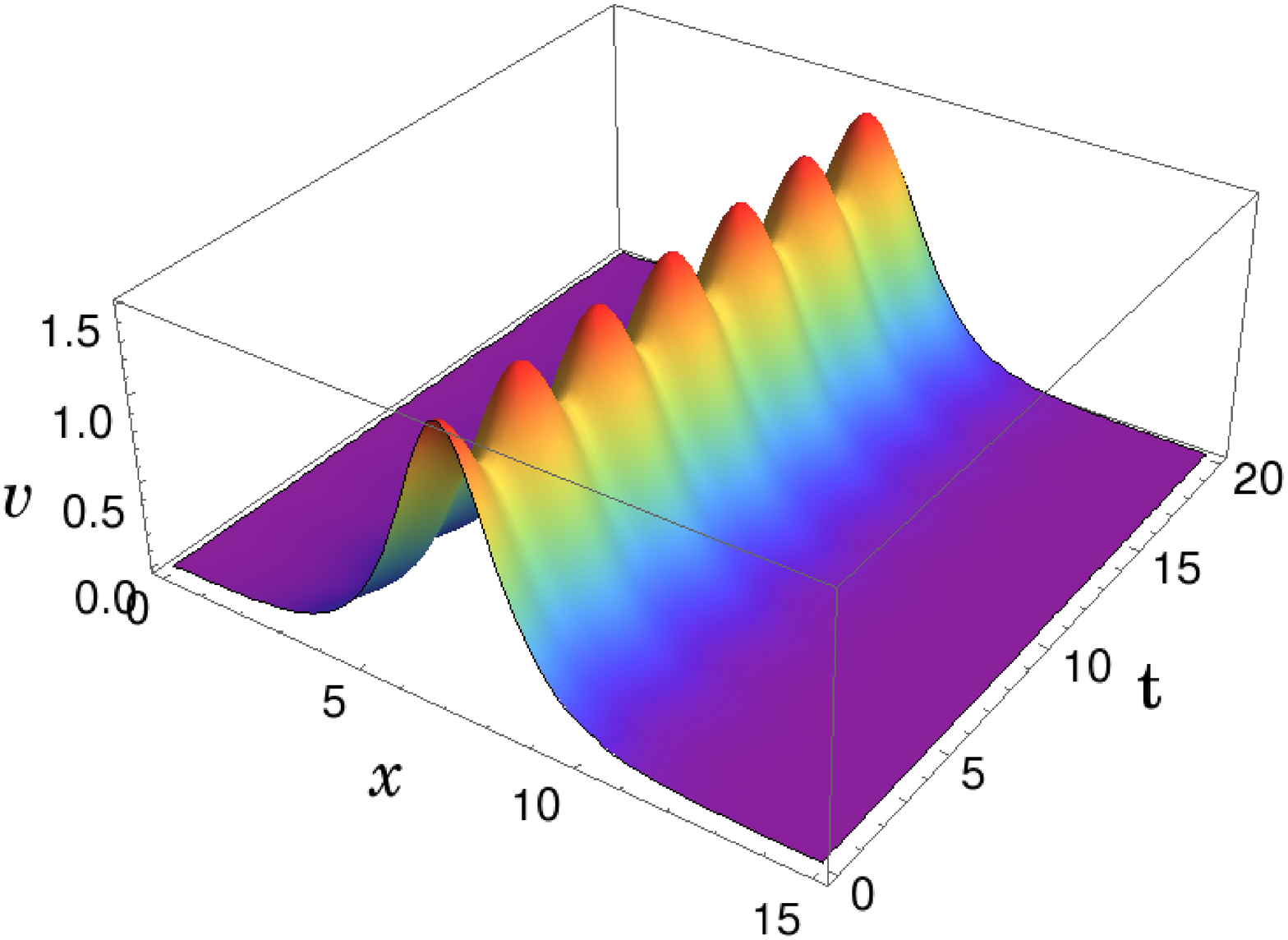}
\caption{(Top panel) Solitary wave solutions of Eq. (\ref{11nls}) for (a) $a_1=1.5$ and (b) $a_1=1.0$ with other parameters as $\gamma=2$, $b=1$, and $k_2=5$. (Bottom panel) The stationary solitary waves with periodically varying amplitude of NLS equation (\ref{nls1}) for the choice of (a) with $y=x$. }
\label{nls-fg2}
\end{figure}
If $c=0$, then the symmetry generator $X_1+b X_3$ reduces
the system of PDEs (\ref{nls1}) to the system of nonlinear
second-order ODEs
\bes\bny
A^{''}-b A+\g A(A^2+B^2)=0, \\
B^{''}-b B+\g B(A^2+B^2)=0.\eny  \label{11nls}\ees
Note that the above system of ODEs (\ref{11nls}) can be viewed as a special case of Eq. (\ref{7}) for $k=a=0$, which is also a Painlev\'e integrable. Interestingly, Eq. (\ref{11nls}) admits the following solitary wave solutions:
\bes\bea
&&A=a_1~ \mbox{sech}(k_1 y+k_2),\\
&&B=a_2~ \mbox{sech}(k_1 y+k_2),
\eea \ees
with the constraints $k_1^2={b}$ and $a_2^2={\frac{2b}{\g}-a_1^2}$. This is shown in Fig. \ref{nls-fg2} for different choices of arbitrary constants.

\begin{figure}[h]
\centering \includegraphics[width=0.35\linewidth]{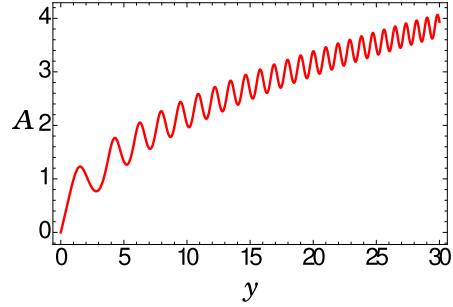}
\caption{Exponentially growing periodic wave trains of Eq. (\ref{12nls}) for $c=1$ and $\gamma=2$.}
\label{fg-12}
\end{figure}

\noindent{\bf Case (v)}: We obtain the group-invariant solution
$u=A(y)\sin(tx/c-2t^3/3c^2)$ and $v=-A(y)\cos(tx/c-2t^3/3c^2)$
from generator $X_4+c X_1$ along with an invariant as $y=c x-t^2$.
This symmetry invariant reduces Eq. (\ref{nls1}) to a nonlinear
second-order ODE of the form
\bny A^{''}-\frac{1}{c^4}yA+\frac{\g}{c^2} A^3=0. \label{12nls}\eny
%where `prime' denotes differentiation with respect to $y$.
We find that the above equation is also Painlev\'e integrable.
The direct numerical analysis shows that the explicit appearance of `$y$' in the model equation induces the modulation of the periodic nonlinear wave structure as it progresses and it can be evidenced from Fig. \ref{fg-12}.

\noindent{\bf Case (vi)}: The symmetry generator $X_3$ does not give any group-invariant solutions.

\section{Conclusions}
We have investigated the symmetry reductions of the nonlinear Helmholtz equation arising in the context of nonlinear optics by using the Lie symmetry analysis. Especially, with an infinitesimal transformation, we have obtained the symmetry generators (vector fields), identified a set of invariants from the optimal system of one-dimensional subalgebras and symmetry reductions in the form of coupled ODEs. We have studied the integrability property of the reduced ODEs by using the Painlev\'e analysis and constructed explicit first integrals by using the modified Prelle-Singer method. In the study of NLH system, we have also obtained nonlinear periodic and solitary wave solutions for some ODEs resulting for specific symmetry reductions. For the remaining ODEs, we have given a qualitative picture of the solutions by numerically solving the underlying ODEs. We have explored some interesting nonlinear wave structures resulting for different choices of arbitrary parameters. The obtained nonlinear periodic wave solutions can find applications in the context of fiber Bragg gratings, bimodal fibers, etc. Further, we have compared the known Lie symmetry analysis of the nonlinear Schr\"odinger equation with the results of the present nonlinear Helmholtz equation. %for a special case (excluding the non-paraxial effect) and compared their results. 
Our analysis shows the possible symmetry reductions of a scalar nonlinear partial differential equation. Thus our study provides an analytical treatment of the interesting NLH system in view of Lie point symmetry, integrability and invariant solutions. 
%It will be an interesting future work to construct explicit solutions of the several ODEs which are addressed numerically in the present work.
As a future study, the present work will be extended to investigate the Lie symmetry and group invariant solutions of various coupled nonlinear evolution equations arising in different contexts.

\section*{Acknowledgments}
KS is grateful to DST-SERB for the award of a National Post-Doctoral Fellowship (PDF/2016/000547). AGJ would like to express his sincere thanks and gratitude to the Centre for Nonlinear Dynamics, School of Physics, Bharathidasan University, Tiruchirappalli, India, for its warmest hospitality and support during his visit. TK is supported by the DST-SERB research project (EMR/2015/001408). The research work of ML is supported by a NASI Senior Scientist Platinum Jubilee Fellowship (NAS 69/5/2016-17) and DST-SERB Distinguished Fellowship (DO No. SB/DF-09/2017). ML is also supported by the research projects of DAE-NBHM (2/48(5)/2015/NBHM(R.P.)/R\&D II/14127), CSIR India  (03/1331/15/EMR-II) and DST-SERB (EMR/2014/001076).

\appendix
\section{Painlev\'e analysis of the ODE (\ref{5})}\label{sec-pain}	%\setcounter{equation}{0}
It is well known that the Painlev\'e analysis gives evidence for the integrability nature of a given system of ODEs/PDEs through an algorithmic procedure, namely leading order analysis, identification of resonances and finding the existence of sufficient number of arbitrary functions at each resonances called arbitrary analysis. For a detailed algorithm see \cite{pain} and references therein.
In this Appendix, we carry out the Painlev\'e analysis for the ODE (\ref{5}) to study its integrability nature.
We express the dependent variables of real equation (\ref{5}) as generalized Laurent series expansion ($A=\ds\sum_{j=0}^{\infty} a_j\phi^{j+\a}$ and $B=\ds\sum_{j=0}^{\infty} b_j\phi^{j+\b}$, where $\a$ and $\b$ are negative integers yet to be determined) in the neighbourhood of the non-characteristic singular manifold $\phi(y)$.
By terminating the Laurent series upto the zeroth order ($A \approx a_0\phi^{\a}$ and $B \approx b_0\phi^{\b}$), we get the leading order of the ODE (\ref{5}) as $\phi^{-3}$ which results for $\a=\b=-1$. Also, we obtain the leading order equation as $2(1+ka^2)+\g(a_0^2+b_0^2)=0$ as the coefficient of $\phi^{-3}$.

From the generalized Laurent series and by using the above leading order equation, we obtain the characteristic equation as $\g(1+ka^2)^2(j+1)j(j-3)(j-4)=0$. So, the resonances for Eq. (\ref{5}) are obtained as $j=-1,0,3,4$, where the resonance $j=-1$ corresponds to the arbitrariness of the non-characteristic manifold. The arbitrary analysis shows that there exists required number of arbitrary functions at each of these resonances. So, we conclude that the coupled second-order ODE (\ref{5}) is Painlev\'e integrable.

Similar to the above analysis for Eq. (\ref{5}), we have carried out the Painlev\'e analysis for other system of ODEs given in this work to identify their integrability nature. The final result of the analysis is given below the respective equations. %(integrable or non-intgerable)

\section{Modified Prelle-Singer method: Solutions for coupled second order ODEs}\label{app-ePS}
In this Appendix, we briefly present the historical developments of the Prelle-Singer method and the important steps involved in the modified Prelle-Singer method. Prelle and Singer, in 1983, proposed a method for finding the general solutions of first-order ODEs \cite{PS}. This method guarantees the solution of a given first-order ODE if the equation under consideration admits an elementary solution. Later, Duarte {\it et al} modified the technique developed by Prelle and Singer and applied it to second order ODEs \cite{Duarte}. Their approach was based on the conjecture that if an elementary solution exists for the given second order ODE then there exists at least one elementary first integral $I(t,x,\dot x)$ whose derivatives are all rational functions of $t$, $x$ and $\dot x$.
This method was recently extended by Chandrasekar, Senthilvelan and Lakshmanan to second order ODEs \cite{ePS}. They have generalized the theory of Duarte {\it et al} and shown that for the second order ODEs one can isolate even two independent integrals of motion and obtain general solutions explicitly without any integration. This method is known as the modified Prelle-Singer method. The method has been extended by these authors to coupled ODEs and higher order ODEs as well.  %This method is known as the modified PS method and have also been successfully generalized to higher order ODEs too.

Let us briefly discuss the modified PS procedure applicable to second order ODEs \cite{ePS} which we have adopted in this manuscript. We assume that the following general second order ODE,
\begin{eqnarray}
\ddot x=\phi(t, x,\dot x),\label{eqn}
\end{eqnarray}
where the over dot denotes differentiation with respect to time, admits a first integral $I(t, x, \dot x)=$ constant. We know that the total derivative of $I$ vanishes, that is,
\begin{eqnarray}
dI=I_{t}dt+I_{x}dx+I_{\dot x}d\dot x=0,\label{I}
\end{eqnarray}
where the subscript denotes partial differentiation with respect to that variable.
Let us now rewrite Eq. (\ref{I}) in the form $\phi dt-d\dot x=0$
and add a null term $S(t,x,\dot x)\dot x dt$ - $S(t,x,\dot x)dx$.  Then we get
\begin{eqnarray}
(\phi+S\dot x)dt-S dx-d\dot x=0. \label{Sform}
\end{eqnarray}
Multiplying the above equation by a factor $R(t, x, \dot x)$, which acts as the integrating factor, Eq. (\ref{Sform}) becomes an exact equation of the form $R(\phi+S\dot x)dt-RSdx-Rd\dot x=0$.
Now, comparing this equation with Eq. (\ref{I}) we get
\begin{eqnarray}
I_{t}=R(\phi+\dot x S),\qquad
I_{x}=-RS, \qquad
I_{\dot x}=-R. \label{IEqs}
\end{eqnarray}
Therefore, the first integral $I$ of Eq. (\ref{eqn}) can be obtained by integrating the above system of equations (\ref{IEqs}) as
\begin{eqnarray}
I=r_{1}-r_{2}-\int{\bigg[R+\frac{d}{d\dot x}(r_{1}-r_{2})  \bigg]d\dot x}, \label{PSI}
\end{eqnarray}
where $r_{1}=\int{R(\phi+\dot x S)dt}, \quad r_{2}=\int{\bigg(RS+\frac{d}{dx}r_{1}  \bigg)dx}$.
Note that the first integral $I(t, x, \dot x)$ is given in terms of $R$ and $S$ which have to be determined. In order to find the explicit forms of $R$ and $S$ for a given form of $\phi$, we use the compatibility conditions ($I_{tx} = I_{xt}$, $I_{t\dot x} = I_{\dot xt}$, and $I_{x\dot x} = I_{\dot xx}$) and the following determining equations for $R$ and $S$:
\begin{subequations}
\begin{eqnarray}
&&S_{t}+\dot x S_{x}+\phi S_{\dot x}=-\phi_{x}+\phi_{\dot x}S+S^2,\\
&&R_{t}+\dot x R_{s}+\phi R_{\dot x}=-(\phi_{\dot x}+S)R, \\
&&R_{x}-S R_{\dot x}-RS_{\dot x}=0.
\end{eqnarray} \label{3c}
\end{subequations}
We would like to emphasis here that any two independent particular solutions $(S, R)$ of the above system of PDEs (\ref{3c}) are enough to deduce two independent first integrals of Eq. (\ref{eqn}). In order to obtain a particular solution of (\ref{3c}), one can use suitable ansatz for $S$ and $R$ and find a compatible solution. Chandrasekar, Senthilvelan and Lakshmanan have also developed a straightforward procedure to deduce the time independent conservative Hamiltonian structure from the time independent first integral obtained by the modified PS procedure \cite{ePS}. Using this procedure the Liouville sense of integrability can be established for the systems for which one is unable to explicitly obtain the second integral of motion.

\end{document}